\documentclass[journal,twoside]{IEEEtran}
\usepackage{float}
\usepackage{graphicx}
\usepackage{color}
\usepackage[square, comma, sort&compress, numbers]{natbib}
\usepackage{hyperref}
\usepackage{subfigure}
\usepackage{amsmath}
\usepackage{bm}
\usepackage{flushend}
\usepackage{algorithm}
\usepackage{algorithmic}
\usepackage{multicol}
\usepackage{booktabs}
\usepackage{makecell}
\usepackage{colortbl}
\usepackage{array,booktabs}

\usepackage{amssymb}
\usepackage{threeparttable}

\usepackage{multicol}
\definecolor{mygray}{gray}{.9}
  \newlength{\halfpagewidth}
    \divide\halfpagewidth by 2

\newtheorem{Rem}{Remark}
\newtheorem{Des}{Question Forming}

\begin{document}

\title{Synthesizing a Clock Signal with Reactions---\\Part II: Frequency Alteration Based on Gears}
\author{Chuan~Zhang,~\IEEEmembership{Member,~IEEE},
        Lulu~Ge,~\IEEEmembership{Student Member,~IEEE},
        Xiaohu~You,~\IEEEmembership{Fellow,~IEEE}\\% <-this % stops a space
\thanks{Chuan Zhang, Lulu Ge, and Xiaohu You are all with National Mobile Communications Research Laboratory, Southeast University, Nanjing, China. Email: \{chzhang, luluge, xhyu\}@seu.edu.cn. Chuan Zhang and Lulu Ge contributed equally to this work. \emph{(Corresponding author: Chuan Zhang.)}}
%\thanks{This paper was presented in part at IEEE Workshop on Signal Processing Systems (SiPS), 2015. \emph{(Corresponding author: Lulu Ge.)}}
}

\markboth{IEEE transactions on xxx,~2017}%
{Chuan Zhang \MakeLowercase{\textit{et al.}}: Synthesizing a Clock Signal with Reactions---Part II: Frequency Alteration Based on Gears}
\maketitle

\begin{abstract}
%\boldmath
On a chassis of gear model, we have offered a quantitative description for our method to synthesize a chemical clock signal with various duty cycles in Part I. As Part II of the study, this paper devotes itself in proposing a design methodology to handle frequency alteration issues for the chemical clock, including both frequency division and frequency multiplication. Several interesting examples are provided for a better explanation of our contribution. All the simulation results verify and validate the correctness and efficiency of our proposal.
\end{abstract}


\IEEEpeerreviewmaketitle

\section{Introduction}\label{sec:1}
%\IEEEPARstart{H}{arnessing} engineering principles to design biological components opens a new era of synthetic biology \cite{synbio}. Indeed, since synthetic biology gained its more modern usage in 1974, a tremendous amount of achievements have been started---synthetic DNA \cite{kosuri2014large,blight2000efficient,gibson2008complete,luo2000synthetic}, designed proteins \cite{broglia1998folding,broglia1999stability}, information storage \cite{church2012next,eigen1966chemical,black1987biochemistry}, biosensors \cite{turner1987biosensors,sassolas2008dna,turner2000biosensors} and even synthetic life \cite{ray1993evolutionary,langton1997artificial,malyshev2014semi}. To date, exciting and novel ideas in this nascent field show its amazing  vigorousness and endless potential. Both biology and engineering call for clock signals to coordinate tasks properly, however. While systematic methods used for designing an electrical clock signal could be available in both hardware and software, but this becomes a bottleneck in synthetic biology.

\IEEEPARstart{C}{lock} signal is essentially an artificial measurement of time, which is of great significance to our life. Under the instruction of clock signals, rhythms abound in biological systems decide the cellular behaviors---cells make decisions and operate assignments from seconds to hours, people work and sleep in a day of $24$ hours, plants blossom and bear fruits in a period of years---underlying those various timing processes, all the rhythms are well-orchestrated. This is the same truth for electrical systems to coordinate tasks. All in all, clock signals undeniably play an important role in both synthetic biology \cite{synbio,andrianantoandro2006synthetic,benner2005synthetic,brenner2008engineering} and electrical systems \cite{sheets1987electrical,kassakian2000automotive,friedman1995clock,russell1996random}.

In Part I of our study, which synthesizes a tunable clock signal in CRN level, we have presented an appropriate gear model to offer: 1) physical analogy for CRN clock design methodology, and 2) a quantitative description of duty-cycle modulation. In other words, with gear models proposed in Part I, our methods to synthesize a chemical clock signal conform to a physical intuition. Unlike Part I, Part II focuses on the frequency alteration instead of duty-cycle modulation. In this paper, by further exploiting the compound gear model, we offer a formal frequency alteration methodology for a clock signal with fixed duty cycle: generating a new clock whose frequency is $L/J$ times of the input clock. Although there exists a limitation on $L$ (being a factor of $N$), the proposed methodology enables us to change the frequency of chemical clock to some extent and makes the frequency processing of CRNs possible. The work shown in Part II is put forward based on the gear theory illustrated in Part I, thus \textcolor{black}{being} considered as a continuation of Part I.

In this paper, under the accumulated instructions derived from Part I, \textcolor{black}{we attempt to address the frequency alteration in two steps: 1) frequency division by $J$, and 2) frequency multiplication by $L$.} Our proposed approaches are validated via numerical simulations of the chemical kinetics based on \textit{ordinary differential equations} (ODEs). It is noted that, the chemical reactions here are all formal chemical reactions, which could be translated into DNA strand displacement reactions if properly designed \cite{soloveichik2010dna}. Therefore, our work owns its physical implementation, namely DNA reactions, although it looks like just pure CRN design.

The rest of this paper is organized as follows. With the definition of fundamental frequency with fixed duty cycle, Section \ref{fd} proposes a method to realize the frequency division by $J$. On a chassis of Part I, a detailed gear model analysis is given in Section \ref{fm} for frequency multiplication by $L$. Different gear combinations for frequency multiplication are taken into consideration in the same section. Concrete examples are offered for a better understanding. Finally, Section \ref{sec:6} concludes the entire paper.

\section{Frequency Division}\label{fd}
%Since frequency division is an important part of frequency alteration,
This section focuses on the frequency division for CRN clocks. %Before proceeding, a fundamental frequency should be defined, because on a chassis of which, frequency division, or more exactly, the frequency alteration could be further conducted. After elaborating our method for frequency division, we also offer a case study for a better explanation.
Frequency alteration is meaningful only after the fundamental frequency (input clock) is defined. More specifically, when it comes to frequency alteration, the compared two clock signals must have the same duty cycle but different frequencies or clock cycles. Theoretically, frequency division implementation tends to address a problem described as follows:
\begin{Des}
\textcolor{black}{For a clock of fundamental frequency $f_{\textrm{in}}$ and a given duty cycle $M/N$, frequency division aims to output a new clock with ${f_{\textrm{in}}}/J$ frequency and unchanged duty cycle. The main elements for this question are listed in Table \ref{tl:1}, from which, only two clock signals are concerned: the fundamental (input) clock signal and the output signal with divided frequency.}
\end{Des}
%\textcolor{red}{\textbf{Design Target.} For a better understanding of our work, here explicitly form the problem that this Part II intends to address. Only after defining the fundamental frequency, which is synthesized with the proposed method in Part I, frequency alteration in Part II would be meaningful. To be more specific, when it comes to frequency alteration, the compared two clock signals must have the same duty cycle but different frequencies or clock cycles. In this sense, the design target can be described as that: for a given input clock signal with $1/N$ duty cycle, we want to obtain its corresponding output signal of frequency alteration. Intuitively, frequency division prolongs the input clock cycle, while frequency multiplication does the opposite. Without loss of generality, we only consider the clock signal frequency alteration with ``1'' as the numerator of its clock duty cycle, namely $1/N$ ($N\!\!>\!\!2$) duty cycle.}
\begin{table}[htbp]
\centering
\caption{Question forming for frequency division}
\begin{tabular}{c||c}
%\begin{threeparttable}
\Xhline{1.0pt}
\textbf{Input} &  \textbf{Output}\\
%\hline
\hline
\rowcolor{mygray}
Fundamental frequency $f_{\textrm{in}}$ &  Divided frequency $f_{\textrm{in}}/J$ \\
%\hline
%Number of gear teeth & Number of phase nodes  \\
%%\hline
%\rowcolor{mygray}
%Diameter of a gear  & Length of a clock period \\
%%\hline
%Angular velocity & \(2\pi\)/clock period\\
\Xhline{1.0pt}
%\hline
%\hline
\textbf{References} &  \textbf{Unchanged parameter}\\
\hline
\rowcolor{mygray}
$N$ and $JN$-phase gears$^*$ & $M/N$ duty cycle\\
%\hline
%Rotation period & Time period \\
%%\hline
%\rowcolor{mygray}
%Color distribution of the rack & Duty cycle\\
\Xhline{1.0pt}
\end{tabular}
\label{tl:1}
%\small Note: Robust standard errors in parentheses. Intercept included but not reported.
\begin{tablenotes}
  \item[1] $^*$Note: $J$ is an integer.
\end{tablenotes}
%\footnotetext{ddd}
%\tablefootnote{ni zhi dao me }
%\end{threeparttable}
\vspace*{-4pt}
\end{table}

\subsection{Fundamental Frequency}\label{p1}
%From our study in Part I, we know that chemical clock signals synthesized with our methods share the same exiting time period \(\hat \tau _{1{\rm{/}}N}\) for each phase when they adopt the same rate constant scheme, where $N \geq 3$. Validated in Part I, through the tool \textit{Matlab}, the rate constant scheme illustrated in Appendix of Part I will make \(\hat \tau _{1{\rm{/}}N}\) to be $0.091$.
Frequency alteration is given based on an implementation of fundamental frequency. For better explanation, here the fixed duty cycle is set as $1/N$. Other duty cycles can be processed in the same fashion. To synthesize a clock signal with CRNs, \(\hat \tau _{1{\rm{/}}N}\) would be $0.091$ when we adopt the rate constant scheme in Appendix of Part I. Therefore, the whole time period of $1/N$ depends on the number of phases the oscillator has, and the exact value of $T_{1{\rm{/}}N}$ is \(N \times \hat \tau _{1{\rm{/}}N}\), namely $0.091N$. Therefore, in our proposal for a $1/N$ ($N\!\!> \!\! 2$) duty cycle clock signal, the fundamental frequency is defined with our previous method.

\begin{figure}[htbp]
\centering
\includegraphics[width=0.75\linewidth]{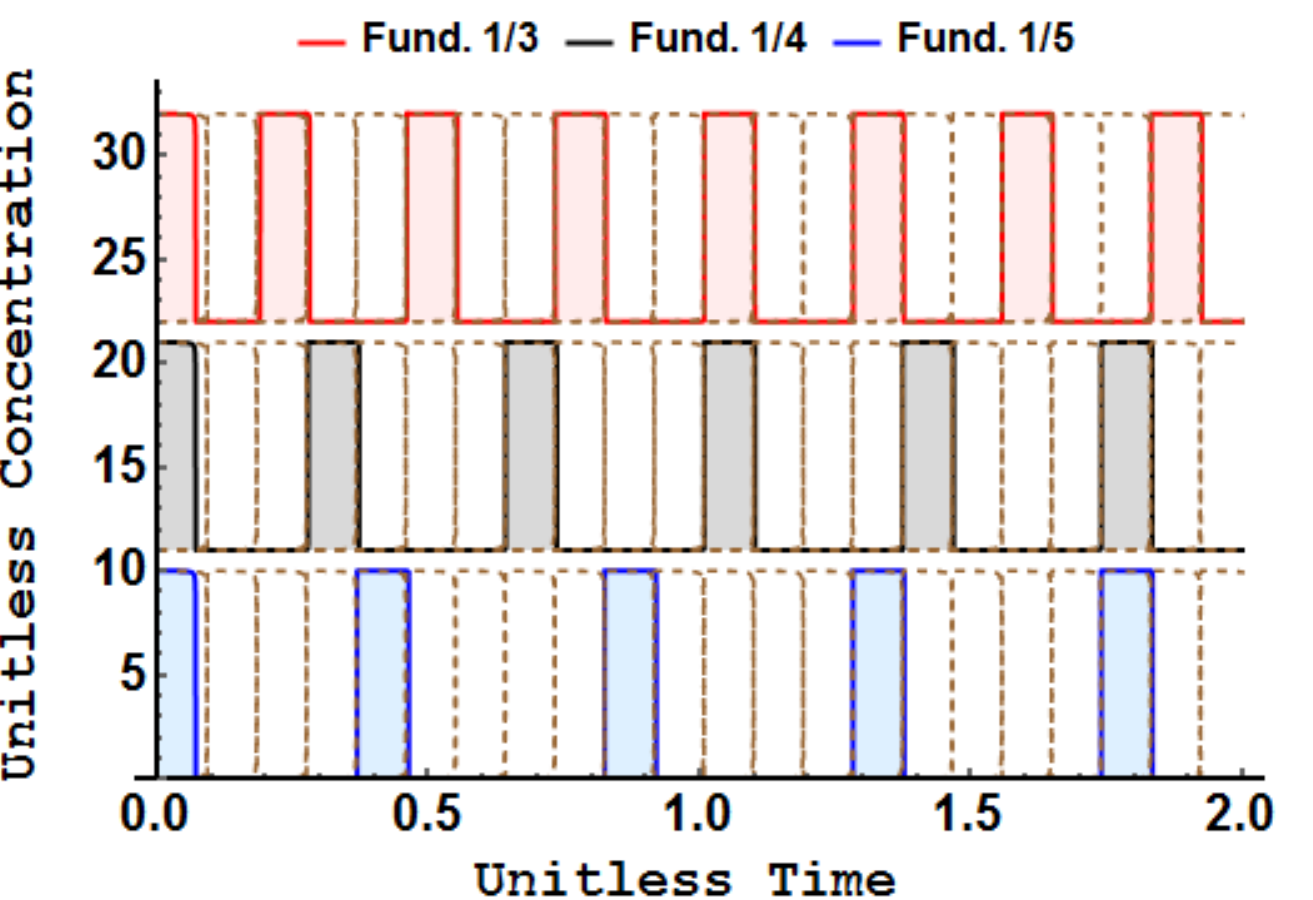}
\caption{Simulation results of $1/3$, $1/4$, and $1/5$ fundamental frequencies.}\label{fig10}
\end{figure}

Fundamental frequencies of  $1/3 $, $ 1/4 $, and $ 1/5 $ duty cycle clocks are shown in Fig. \ref{fig10}. These results show that they have different time period $T_{1{\rm{/}}N}$ but the same phase existing time \(\hat \tau _{1{\rm{/}}N}\). One thing should be emphasized is that, the fundamental frequencies might have a clock skew at the first beginning. And the reason for this phenomenon is the first touch between two meshed gears. After this first ``uncomfortable'' touch, everything will be okay, including the oscillation and the produced clock signals.

%\begin{Thm}\label{thm:3}
%The existing period of each phase signal \(\tau _{1{\rm{/}}N} \approx 0.091\) is the standard length of a clock signal in CRNs.
%\end{Thm}

%\begin{IEEEproof}
%Essentially, the clock signal of $1/N$ duty cycle works with the primitive oscillation offered by an $N$-phase oscillator, implemented with our previous method. On a chassis of the same oscillating principle, the existing period of each phase signal is identical. Simulation results in Fig. \ref{fig10} further validate this point.
%\end{IEEEproof}

Hence in our proposal, the fundamental frequency is defined as those clock signals implemented with our previous method for $1/N$ $(N>2)$ duty cycle, or rather the time period measured by \(\hat \tau _{1{\rm{/}}N}\). The corresponding gear model is the implementation of $1/N$ duty cycle illustrated in Section VI of Part I. Owing to this fundamental frequency, frequency alteration could be further conducted. Additionally, frequency division could be easily understood as it has a longer time period while frequency multiplication has a shorter one.

\subsection{Rationale of Frequency Division}\label{p2}
In traditional electronics, for a given fundamental frequency $f_{\textrm{in}}$, the realization of ${f_{\textrm{in}}}/J$ indicates frequency division, where $J$ is an integer. In our proposal for CRNs, $J$ multiple for both numerator and denominator of ${f_{\textrm{in}}}$ gives frequency division, where the fundamental frequency is defined based on the standard implementation of $1/N$ duty cycle clock signal. For instance, on a chassis of a $1/N$ duty cycle clock, frequency division could be realized through the construction of $2/2N$, $3/3N$, $ \cdots $, and $n/nN$.

\subsection{Case Study for Frequency Division}\label{sec:pp1}
\begin{figure}[htbp]
\centering
\includegraphics[width=0.75\linewidth]{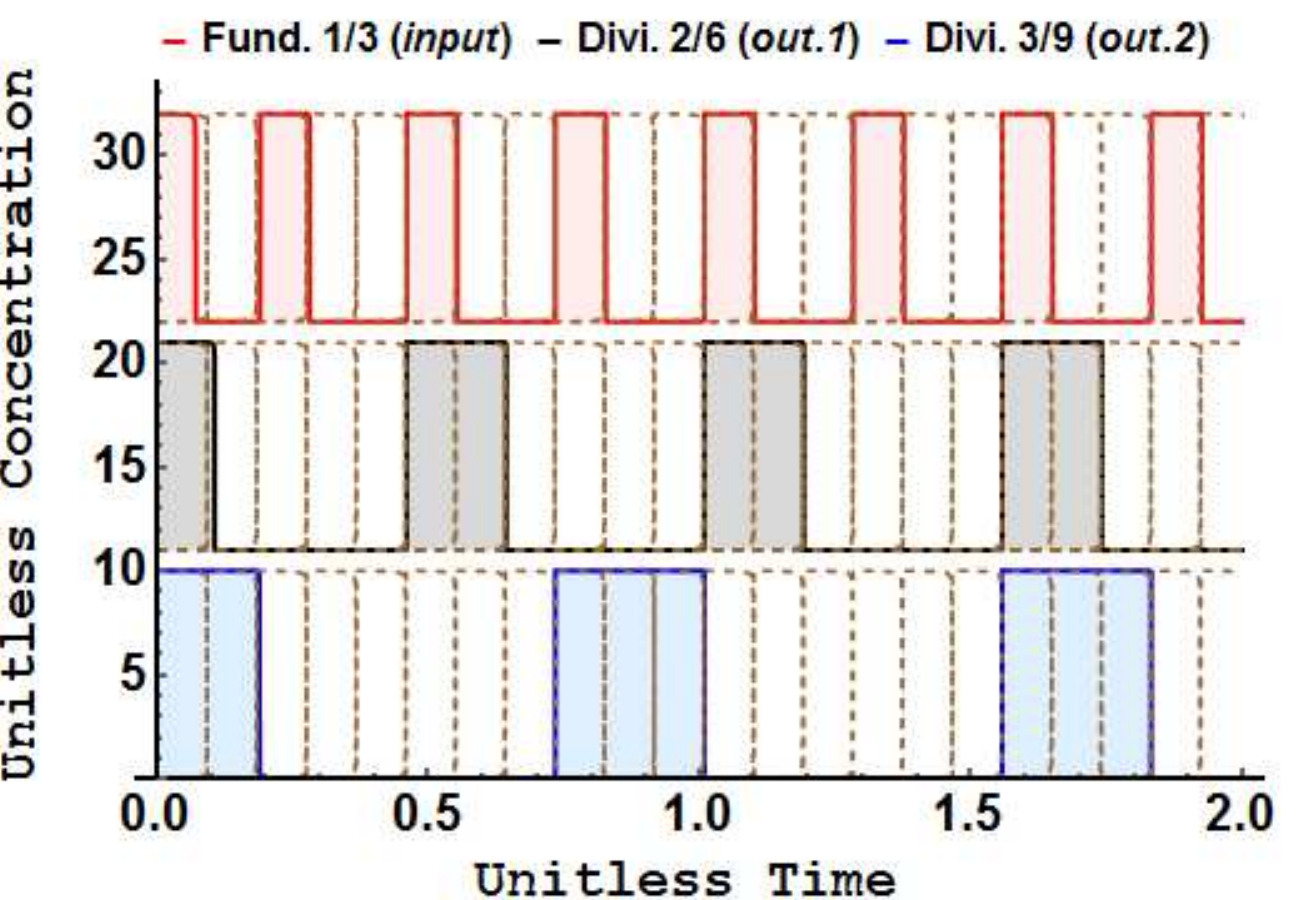}
\caption{Simulation of $1/3$ fundamental frequency and frequency division.}\label{figfdi}
\end{figure}

Take a $1/3$ duty cycle clock signal as an example. As shown in Fig. \ref{figfdi}, the fundamental frequency of $1/3$ duty cycle is the top red curve, which is synthesized with our method illustrated in Part I. Its frequency division could be realized by implementing clock signals with duty cycles of $2/6$, $3/9$, and so on. The other two curves in Fig. \ref{figfdi}, representing $2/6$ and $3/9$ respectively, realize the wanted frequency divisions. Both $2/6$ and $3/9$ duty cycle clock signals are constructed according to the implementation methodology of $M/N$ duty cycle clock signal in Part I. Each phase signal is colored brown by a dashed line in this figure.

%\begin{figure}[htbp]
%\centering
%\includegraphics[width=0.8\linewidth]{15divi.pdf}
%\caption{Simulation of $1/15$ fundamental frequency. Both the initial concentration of $clk_{1}$ and $R$ are $10$ unitless density. The layout of $clk_{1}$ ranges from $11$ to $21$ in the perspective of its concentration.}\label{fig9}
%\end{figure}

%Frequency division could be realized following the specific steps explained in Algorithm \ref{arg1}. Chemical reactions with rate constants are also given in Eq. \ref{eq1:1} in Appendix \ref{sec:app}. We use a $15$-phase oscillator to mesh the gear of $1/2$ duty cycle. Each phase ranges from $R$, $R_{1}$, ..., $R_{14}$. Fig. \ref{fig9} shows the corresponding simulation result.

%Use phase signal $R$ and $R_{14}$ of this $15$-phase oscillator to control the transference between $clk_{0}$ and $clk_{1}$. From Fig. \ref{fig9}, especially the phase signal $R$ filled with light blue, \textcolor{red}{no change triggers in the signal of $15$-phase oscillator but exists in the final output signal.} Therefore, Eq. \ref{eq1:1} realizes $1/15$ duty cycle. Actually in Fig. \ref{fig9}, a clock skew occurs in the beginning of $clk_{1}$. From a perspective of gear system, this result from the mesh between $G_{1}$ of $1/2$ and $G_{3}$ of $1/15$. After the very beginning mesh, the mechanism works well \textcolor{red}{(figure)}.

Hence, the clock signal synthesized with $1/N$ ($N > 2$) duty cycle in Part I would be viewed as a fundamental frequency. Frequency division would be realized through the implementation of $M/N$ duty cycle, which has its physical meaning of longer time period than the fundamental one. The corresponding gear models can also be found in Part I. %\textcolor{red}{Gear model for frequency division}.

\section{Frequency Multiplication}\label{fm}
Based on the related work in Part I for compound gears, this section focuses on the frequency multiplication issue. Two methods are given in this section. Frequency multiplication of CRN clock aims to address the following problem:
\begin{Des}
\textcolor{black}{For a clock of fundamental frequency $f_{\textrm{in}}$ and a given duty cycle $M/N$, frequency mulitiplication aims to output a new clock with $L{f_{\textrm{in}}}$ frequency and unchanged duty cycle. The main elements for this question are listed in Table \ref{tl:2}, from which, only two clock signals are concerned: the fundamental (input) clock signal and the output signal with divided frequency.}
\end{Des}
\begin{table}[htbp]
\centering
\caption{Question forming for frequency multiplication}
\begin{tabular}{c||c}
%\begin{threeparttable}
\Xhline{1.0pt}
\textbf{Input} &  \textbf{Output}\\
%\hline
\hline
\rowcolor{mygray}
Fundamental frequency $f_{\textrm{in}}$ &  Multiplied frequency $L{f_{\textrm{in}}}$ \\
\Xhline{1.0pt}
\textbf{Reference signal} &  \textbf{Unchanged parameter}\\
\hline
\rowcolor{mygray}
$K$ and $L$-phase gears$^*$ & $1/N$ duty cycle\\
%\hline
%Rotation period & Time period \\
%%\hline
%\rowcolor{mygray}
%Color distribution of the rack & Duty cycle\\
\Xhline{1.0pt}
\end{tabular}
\label{tl:2}
%\small Note: Robust standard errors in parentheses. Intercept included but not reported.
\begin{tablenotes}
  \item[1] $^*$Note: $N=K \times L$.
\end{tablenotes}
%\footnotetext{ddd}
%\tablefootnote{ni zhi dao me }
%\end{threeparttable}
\vspace*{4pt}
\end{table}

\subsection{Rationale of Frequency Multiplication}\label{bi}
\begin{figure}[htbp]
\centering
\includegraphics[width=0.6\linewidth]{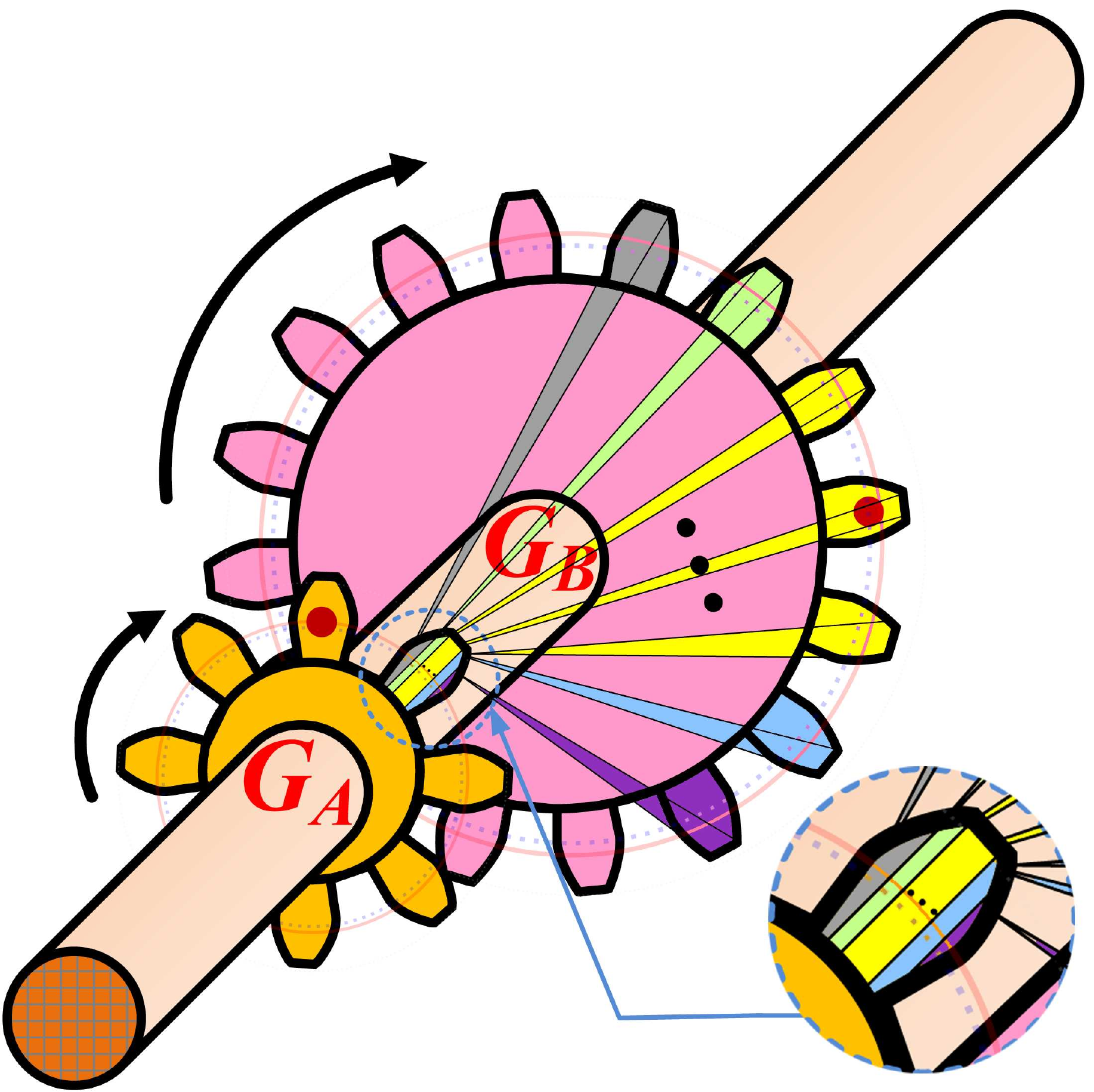}
\caption{A coarse gear model for frequency multiplication.}\label{fig6}
\end{figure}

Our frequency multiplication is actually based on the compound gears illustrated in Part I. The more specific design inspiration is derived from Part I by segmenting a phase signal into pieces and adopting an appropriate rate constant adjustment. For a better explanation of our proposal, a detailed illustration is given as follows.
\begin{figure}[htbp]
\centering
\includegraphics[width=0.8\linewidth]{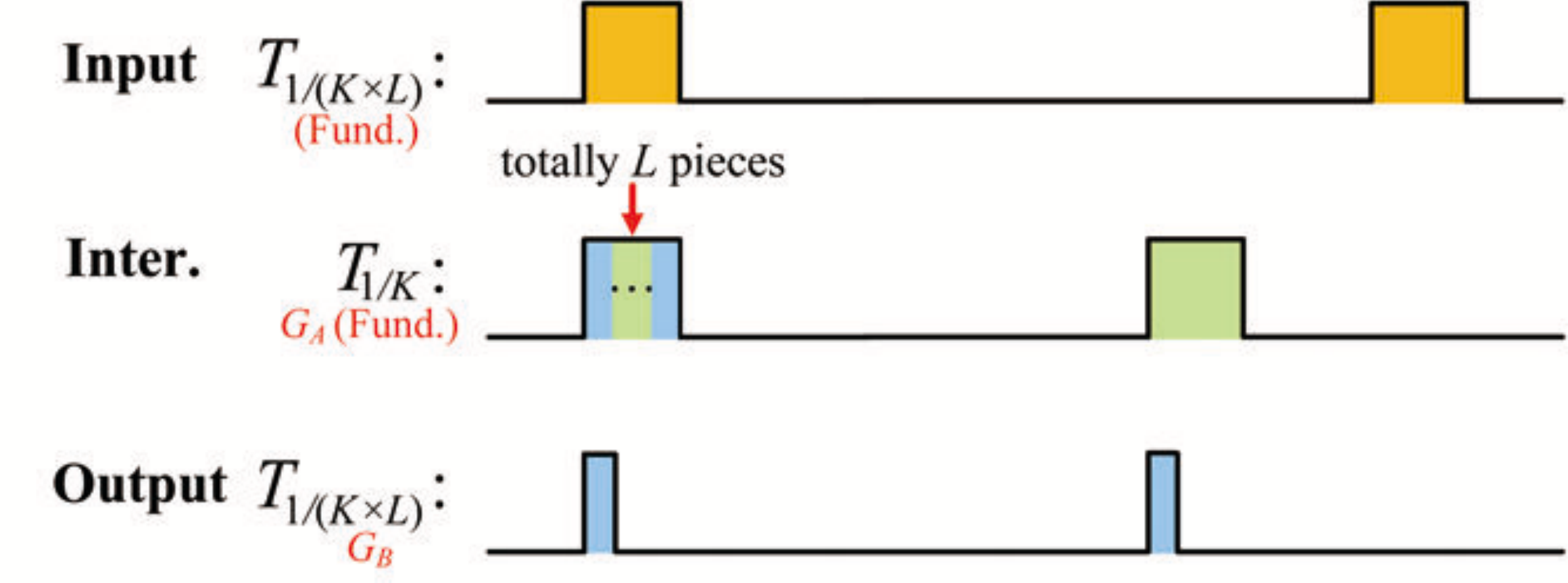}
\caption{Clock signals of $1/K$ and $1/(K \times L)$, where $K \times L =N$.}\label{fig7}
\end{figure}

First of all, a compound gear model is shown as Fig. \ref{fig6}. Assume $G_A$ is a $K$-phase oscillator, and $G_B$ is an $N$-phase one. Additionally, if one tooth of $G_A$ denotes a phase signal of the $K$-phase oscillator, the enlarged colorful painting scheme is actually to equally divide this phase signal's existing time into $L$ pieces, and each piece corresponds to a phase signal of $G_{B}$. Moreover, this segmentation essentially realizes a $1/N$ duty cycle, that is $G_B$, where $N=K \times L$. The corresponding clock signals for both $G_A$ and $G_B$ are shown in Fig. \ref{fig7}.

\textcolor{black}{From Fig. \ref{fig7}, the input clock signal of the design target in Section \ref{sec:1} is the top one, which is a fundamental frequency of $1/N$ duty cycle. The remained two signals are the results of our compound gear model shown in Fig. \ref{fig6}. The $1/K$ duty cycle signal (denoted by $G_A$) has the same pulse width with the fundamental frequency $1/N$ one, their difference is the teeth number. Additionally, this $1/K$ duty cycle signal owns the same time period of the produced $1/N$ one (represented by $G_B$), which has shorter time period of its fundamental frequency. Thus, this operation based on the compound gear model does really realize frequency multiplication. The target clock signal for frequency multiplication, represented by $G_{B}$, could be obtained via three steps summarized in Algorithm \ref{h}. Note that the number of these chemical reactions is less than that of reactions realizing the same duty cycle clock signal with our method in Part I.}

\begin{Rem}\label{rem8}
The property of this frequency multiplication could be realized by ``phase signal controlling''. It means only one phase signal of $K$-phase oscillator controls the transference of $L$-phase oscillator, especially the process of threshold and main power reactions. Meanwhile, two phase signals of $L$-phase oscillator are used to control the whole transference of $1/2$ duty cycle clock signal.
\end{Rem}

\begin{algorithm}
\caption{CRNs to implement $G_{B}$.}\label{h}
\begin{algorithmic}[1]
\REQUIRE CRNs for $1/K$, $1/L$ and $1/2$ duty cycle.
\STATE Construct a $K$ and $L$-phase oscillator with CRNs, respectively, where \(K,L \ge 3\).
\STATE Implement a $1/2$ duty cycle with $12$ chemical reactions.
\STATE Use one phase signal of $K$-phase oscillator to control the transference of $L$-phase one.
\STATE Use two phase signals of $L$-phase oscillator to control the transference of $1/2$ duty cycle.
\STATE Detect one phase of $1/2$ clock signal, the duty cycle of final clock signal could range from $\frac{1}{K \times L}$, $\frac{2}{K \times L}$, $ \ldots $ $\frac{1}{K}$.
\end{algorithmic}
\end{algorithm}

\begin{Rem}\label{rem9}
Implementing in Algorithm \ref{h} requires a total of $(4K+4L+12)$ chemical reactions. Whereas in our previous work, the same duty cycle calls for $(4N+12)$ reactions, where \(K,L \ge 3\). Hence, the method in Fig. \ref{fig6} could not only realize the frequency multiplication, but also reduce chemical reactions because \(KL \gg (K + L)\).
\end{Rem}

%\subsection{Case study for Frequency Multiplier}

\begin{figure*}[htbp]
\centering
\includegraphics[width=0.8\linewidth]{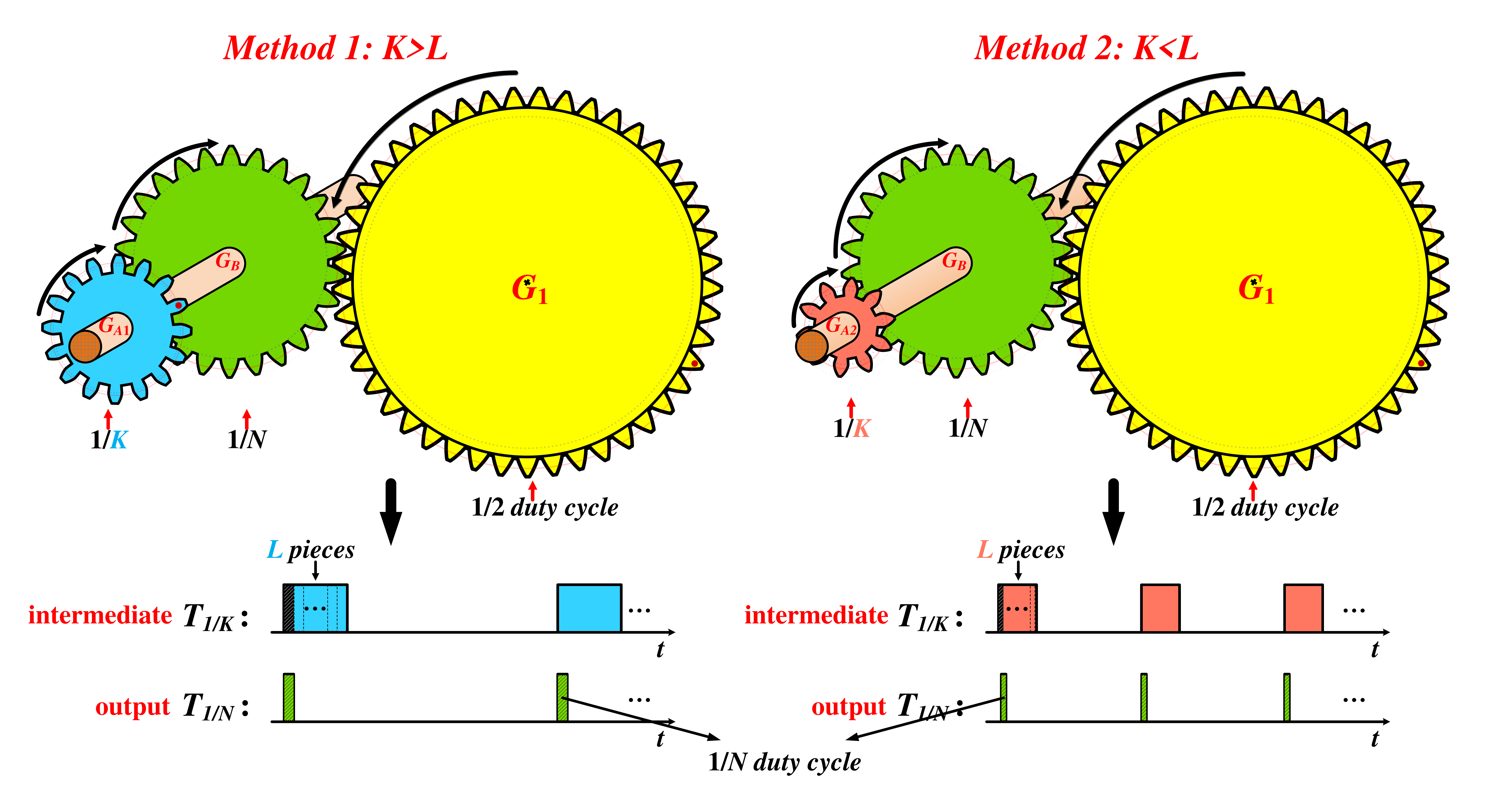}
\caption{Gear model for frequency multiplication with Method 1 and Method 2 when $K \neq L$.}\label{figm}
\end{figure*}

%\subsection{Frequency Division}\label{pp1}
%As shown in Fig. \ref{fig7}, the basic idea proposed in Section \ref{bi} essentially realizes the same duty cycle with our previous method, the existing time period is actually changed shorter, however. More exactly, although $N=KL$, the clock signal of $\frac{1}{K \times L}$ duty cycle has shorter $\tau_{1{\rm{/}}KL}$ than $\tau_{1{\rm{/}}N}$. Thus, both ``Frequency Multiplier'' and ``Frequency Division'' are induced. Without  loss of generality, a fundamental frequency should be defined well. On the basis of this fundamental frequency, the concept of ``Frequency Multiplier'' and ``Frequency Division'' would have physical meaning. This subsection mainly focus on ``Frequency Division''.

%If observe carefully, we could find that the rationale for an implementation of $1/N$ clock signal with fewer reactions illustrated in Section \ref{bi} is exactly an intuition of ``Frequency Multiplier''. This is because the corresponding $\tau_{1{\rm{/}}N}$ of final clock signal is no longer but shorter than the standard $0.091$.
\subsection{Different Conditions for Frequency Multiplication}
Since the aforementioned phase signal controlling is given by segmenting one phase of $K$-phase oscillator into $L$ pieces, conditions are categorized for: $K \neq L$ and $K = L$. For $K \neq  L$, two methods are proposed.

\subsubsection{\textbf{Frequency Multiplication When $K \neq L > 2$}}
On the premise of $K \neq L$, two methods are proposed to address $K>L$ and $K<L$ issues, named Method 1 and Method 2, respectively. Both conditions segment one phase signal of $K$-phase oscillator into $L$ pieces. Method 1 and Method 2 do not only differ in terms of the values of $K$ and $L$, but also the time period of output clock signals, as well as the rate constant adjustment schemes they require. The corresponding gear models are shown in Fig. \ref{figm}.

%Additionally, since the realization of ``Frequency Multiplier'' equals to an implementation of $1/N$ clock signal, which requires two oscillators with respective $K$ and $L$ phases. Hence for the same frequency multiplier of $1/N$ clock signal, there exist two methods depending on whether $K$ controls $L$ oscillator or $L$ controls $K$ oscillator when $K>L$.

%On the premise of $K>L$, to put it simple, if $K$ oscillator controls the transference of $L$ oscillator, the corresponding approach is called \textbf{Method 1}, which uses a big oscillation to control a smaller one. While \textbf{Method 2} is an approach that uses a small oscillation to control a bigger one, in the case of $L$ oscillator controls the $K$ oscillator. The difference between \textbf{Method 1} and \textbf{Method 2} is not only the control order, namely $K$ control $L$ or $L$ control $K$ oscillator, but also a rate constant adjustment.

\begin{figure}[htbp]
\centering
\subfigure[Fundamental frequency and failed multiplication.]{
\label{fig12:a}
\includegraphics[width=0.75\linewidth]{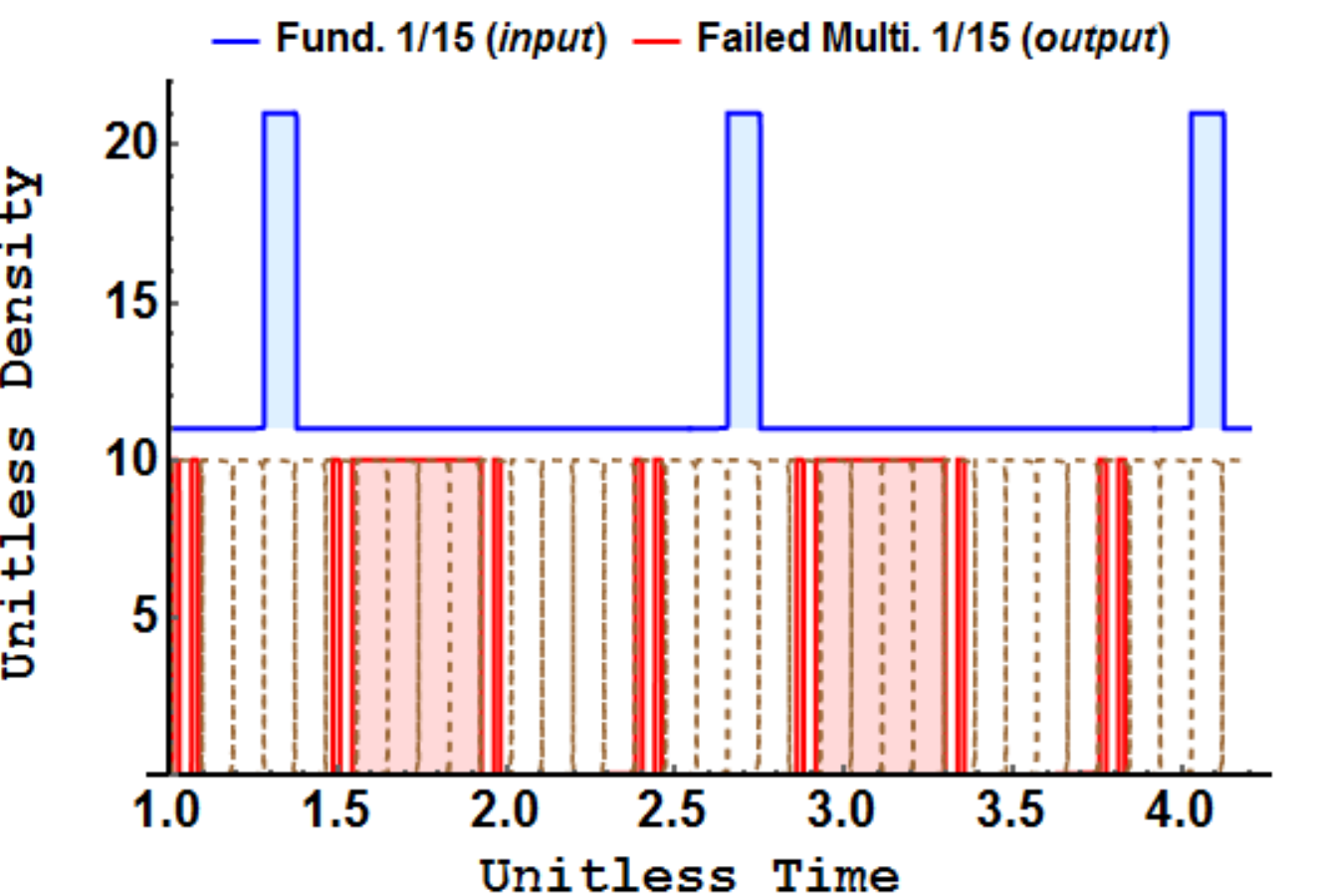}}
\subfigure[Output of $K$ oscillator and frequency multiplication.]{
\label{fig12:b}
\includegraphics[width=0.75\linewidth]{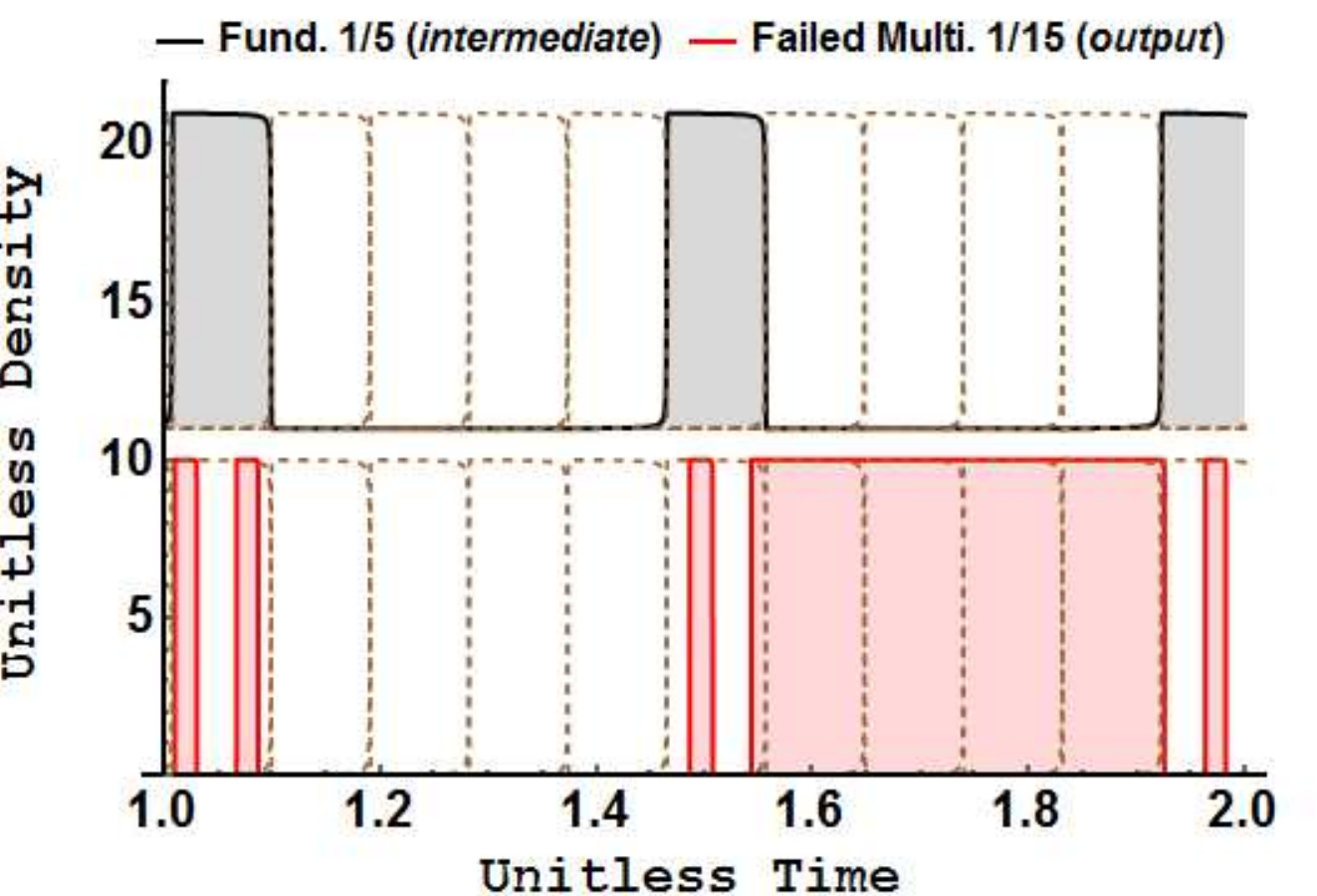}}
%\subfigure[The output of gear system.]{
%\label{fig12:a}
%\includegraphics[width=0.48\linewidth]{xdr1.pdf}}
%\subfigure[Simulation of single RGB.]{
%\label{fig12:b}
%\includegraphics[width=0.48\linewidth]{xdr2.pdf}}
%\subfigure[Simulation of single $G_{A1}$.]{
%\label{fig12:c}
%\includegraphics[width=0.48\linewidth]{xdr3.pdf}}
%\subfigure[Simulation of $K$ and $L$ oscillator.]{
%\label{fig12:d}
%\includegraphics[width=0.48\linewidth]{xdr4.pdf}}
\caption{Simulation results for Method 1 without rate constant adjustment.}\label{fig12}
\end{figure}

\paragraph{\textbf{Method 1 for $K>L$}} Three gears are required in our gear model, namely $G_{A1}$, $G_B$ and $G_1$, representing oscillators of $1/K$, $1/N$ and $1/2$ duty cycle, respectively. The rationale for $G_{A1}$ and $G_B$ has been illustrated in Fig. \ref{fig6}. More specifically, to realize this target $1/N$ duty cycle, we harness one phase signal of CRNs for $1/K$ duty cycle to control the whole transference of CRNs for $1/L$ duty cycle. In practice, we often use one phase of the $K$-phase oscillator to manipulate the threshold and main power reactions of the CRNs for the $L$-phase oscillator. If CRNs of the $K$-phase oscillator adopt the standard rate constant scheme, one thing should be emphasized is that, the rate constant of main power reactions for $L$-phase oscillator should be slowed down. Otherwise, an unwanted oscillation will occur.

\begin{figure}[htbp]
\centering
\subfigure[Fundamental frequency and frequency multiplication.]{
\label{fig13:a}
\includegraphics[width=0.75\linewidth]{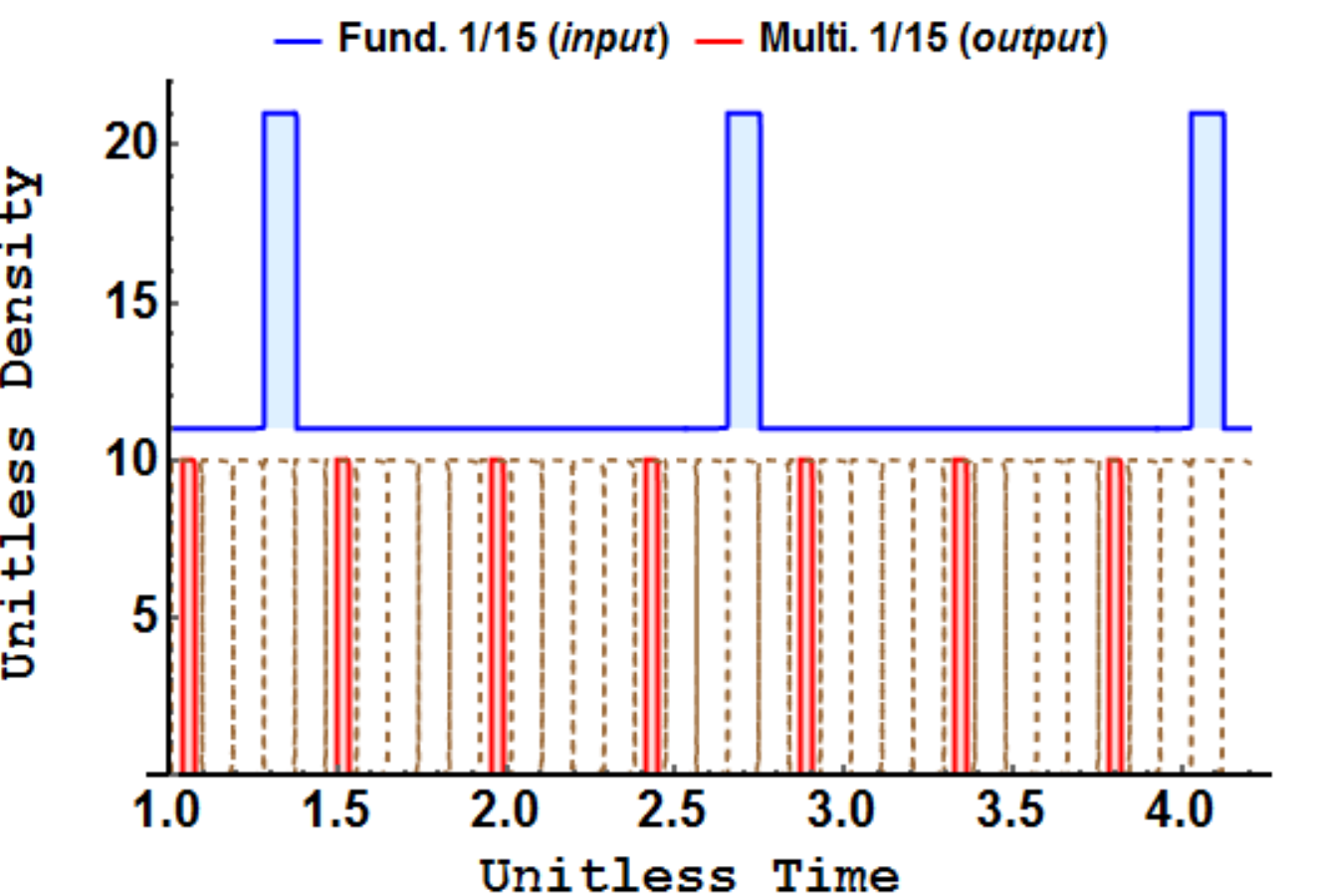}}
\subfigure[Output of $K$ oscillator and frequency multiplication.]{
\label{fig13:b}
\includegraphics[width=0.75\linewidth]{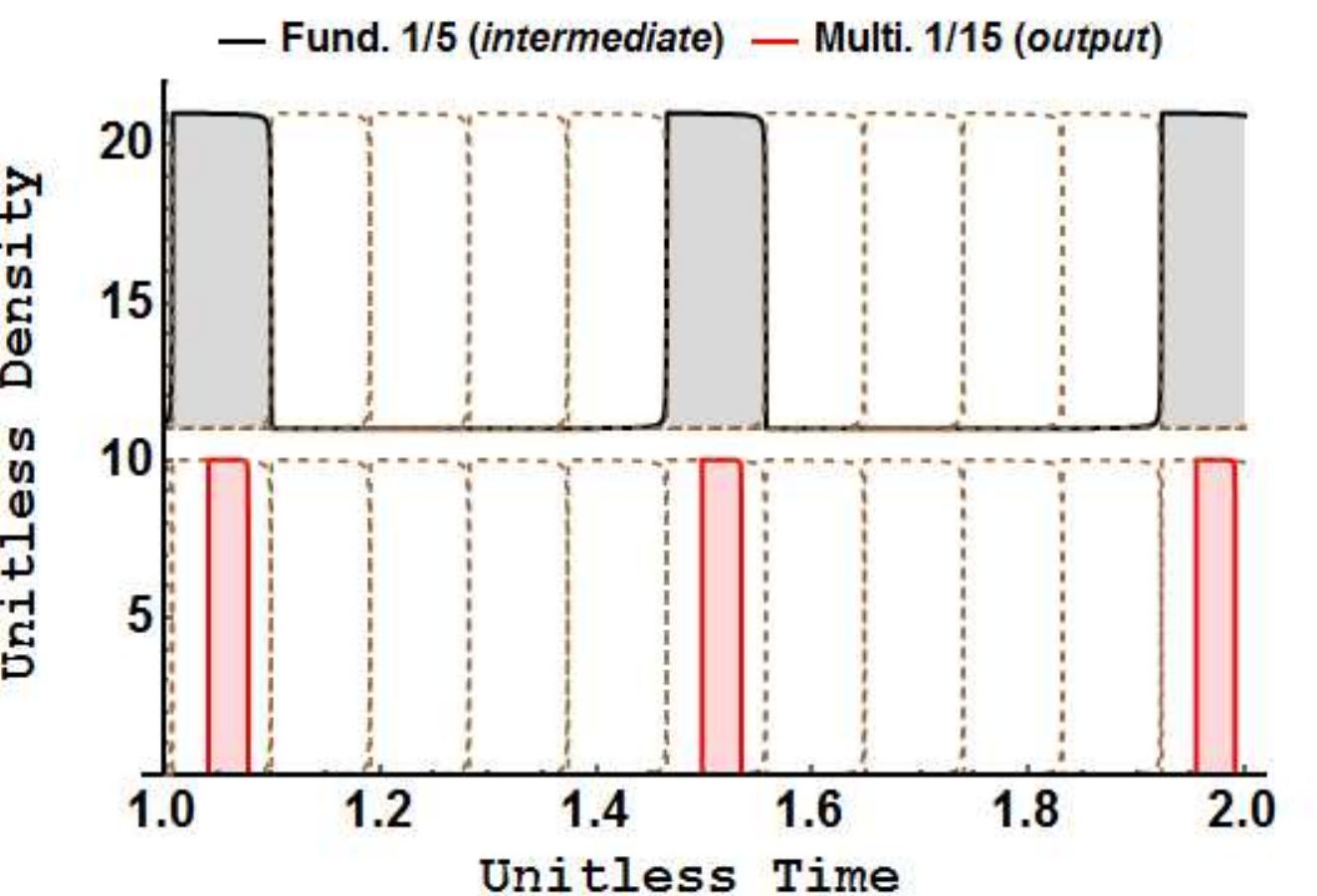}}
%\subfigure[The output of gear system.]{
%\label{fig13:a}
%\includegraphics[width=0.48\linewidth]{xd1.pdf}}
%\subfigure[Simulation of single RGB.]{
%\label{fig13:b}
%\includegraphics[width=0.48\linewidth]{xd2.pdf}}
%\subfigure[Simulation of single $G_{A1}$.]{
%\label{fig13:c}
%\includegraphics[width=0.48\linewidth]{xd3.pdf}}
%\subfigure[Simulation of $K$ and $L$ oscillator.]{
%\label{fig13:d}
%\includegraphics[width=0.48\linewidth]{xd4.pdf}}
\caption{Simulation results for Method 1 with rate constant adjustment.}\label{fig13}
\end{figure}

\qquad \textbf{\textit{Example.}} An example of frequency multiplication with $1/15$ duty cycle is given for our proposal. Since $15\!\!=\!\!3\!\! \times \!\!5$ and $K\!\!>\!\!L$, we have $K\!\!=\!\!5$ and $L\!\!=\!\!3$ in Method 1. Synthesized with the methods in Part I and introduced the phase signal controlling procedure, the final clock signal is shown as the bottom red curve in Fig. \ref{fig12} if no rate constant adjustment has been taken. Note that the red curve in Fig. \ref{fig12:b} is an enlarged version of the red one in Fig. \ref{fig12:a}. From Fig. \ref{fig12}, an unwanted oscillation occurs since all the CRNs for oscillators adopt the standard rate constant scheme. Things will be much better if the rate constant of main power reactions for $L$-phase oscillator is changed from $100$ to $25.5$. The corresponding results are shown in Fig. \ref{fig13}.

%For the convenience of description, $G_{A1}$ of $5$-phase oscillation is denoted as a YMPQF oscillator, while a $3$-phase oscillation is denoted as an RGB one. Therefore, with \textbf{Method 1}, a phase signal of YMPQF oscillator, for example Y, is used as a control signal to control the transference of RGB oscillator. While two phases of RGB oscillator are utilized to control the transference of $G_{1}$. If no rate constant adjustment is adopted in \textbf{Method 1}, the corresponding simulation results are shown in Fig. \ref{fig12}.

\qquad \textbf{\textit{Results and Analysis.}} Verified by Fig.s \ref{fig12} and \ref{fig13}, our $K$-phase oscillator, or rather the five-phase oscillator, operates well and produces the standard fundamental frequency of $1/5$ duty cycle in a black curve colored gray. However, chaos occurs when no appropriate rate constant adjustment is adopted in the CRNs of $L$-phase oscillator. The main reason for this chaos in Fig. \ref{fig12} is the too rapid transference of $L$-phase oscillator.
Because the standard fundamental frequency of $1/3$ duty cycle has shorter time period than that of $1/5$ one, which means the transference rate of $3$-phase oscillator is much faster. Seize this key point, the rate constant of main power reactions for $3$-phase oscillator, under the control of one phase signal of $5$-phase oscillator, should be slowed down. Validated by Fig. \ref{fig13}, the final frequency multiplication of $1/15$ duty cycle colored red in Fig. \ref{fig13:b} works well, since it has a shorter time period than the blue-colored fundamental one in Fig. \ref{fig13:a}.

\paragraph{\textbf{Method 2 for $K<L$}} Similar to Method 1, the gear model for this method still requires three gears shown in Fig. \ref{figm}. They are $G_{A2}$, $G_B$, and $G_1$, representing oscillators of $1/K$, $1/N$, and $1/2$ duty cycles, respectively. The only difference beteen Method 2 and Method 1 is that the rate constant of main power reactions for $L$-phase oscillator should be a little faster. In other words, it should be greater than $100$.

%\begin{figure}[htbp]
%\centering
%\subfigure[The output of gear system.]{
%\label{figg7:a}
%\includegraphics[width=0.48\linewidth]{c1.pdf}}
%\subfigure[Simulation of single \({G_A2}\).]{
%\label{figg7:b}
%\includegraphics[width=0.48\linewidth]{c2.pdf}}
%\subfigure[Simulation of single YMPQF.]{
%\label{figg7:c}
%\includegraphics[width=0.48\linewidth]{c3.pdf}}
%\subfigure[Simulation of  $K$ and $L$ oscillator.]{
%\label{figg7:d}
%\includegraphics[width=0.48\linewidth]{c4.pdf}}
%\caption{Simulation results of Method 2 without rate constant adjustment.}\label{figg7}
%\end{figure}

\begin{figure}[htbp]
\centering
\subfigure[Fundamental frequency and frequency multiplication.]{
\label{fig14:a}
\includegraphics[width=0.75\linewidth]{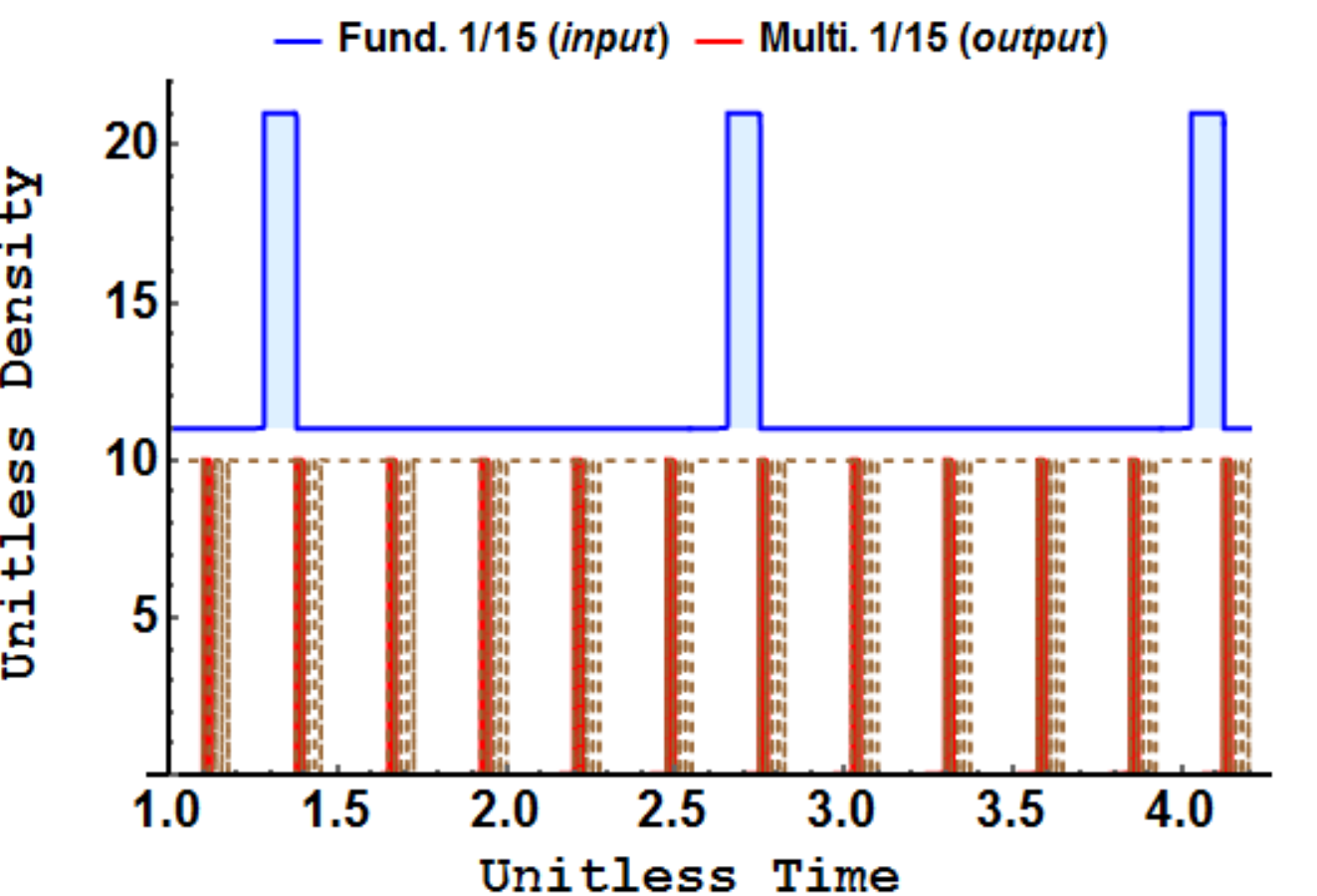}}
\subfigure[Output of $K$ oscillator and frequency multiplication.]{
\label{fig14:b}
\includegraphics[width=0.75\linewidth]{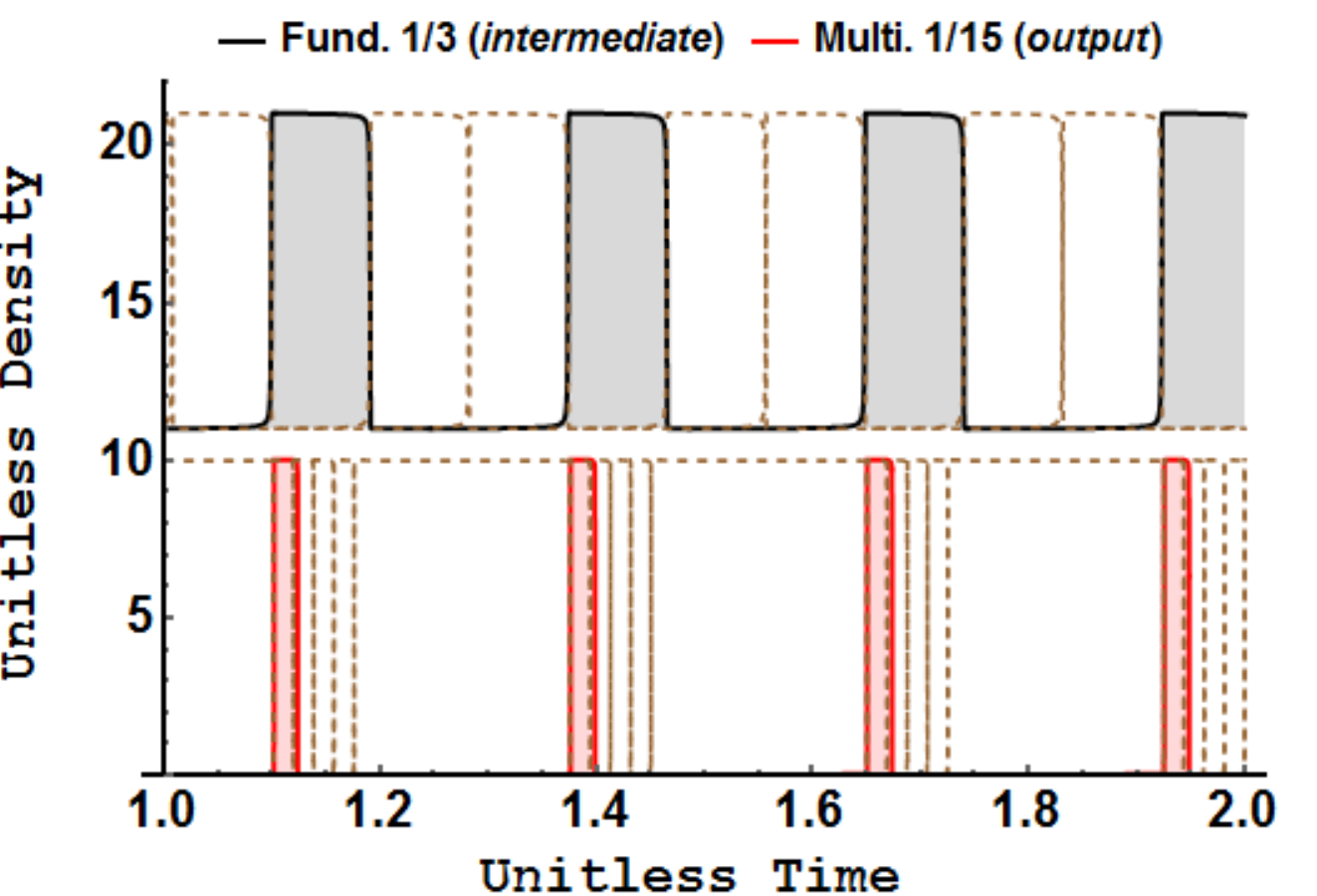}}
%\subfigure[The output of gear system.]{
%\label{fig14:a}
%\includegraphics[width=0.48\linewidth]{dx1.pdf}}
%\subfigure[Simulation of single $G_{2}$.]{
%\label{fig14:b}
%\includegraphics[width=0.48\linewidth]{dx2.pdf}}
%\subfigure[Simulation of single $G_{4}$.]{
%\label{fig14:c}
%\includegraphics[width=0.48\linewidth]{dx3.pdf}}
%\subfigure[Simulation of $K$ and $L$ oscillator.]{
%\label{fig14:d}
%\includegraphics[width=0.48\linewidth]{dx4.pdf}}
\caption{Simulation results of Method 2 with rate constant adjustment.}\label{fig14}
\end{figure}

\qquad \textbf{\textit{Example.}} Still take the $1/15$ duty cycle frequency multiplication as an example. This time $K=3$ and $L=5$. Similar to Method 1, a rate constant adjustment should be adopted in this method. The corresponding results are shown in Fig. \ref{fig14} when the rate constant of main power reactions for $L$-phase oscillator is set as $104$.

%\textbf{\textit{Results and Analysis.}} One thing should be emphasized is that, although the Fig. \ref{figg7:a} merely shows $4$ dashed lines in brown, and each small section is $1/5$ of a phase signal of \({G_{A2}}\). This could be found through Fig. \ref{figg7:d}. Note that an imprecise segment occurs in Fig. \ref{figg7:d}. One efficient solution to this problem is still to change the rate constant of main power reactions. After an analysis, the rate constant should be faster. Because the whole five phases should already oscillate in the existing period of R phase signal, while the whole oscillating period of RGB is shorter than another. In this case, the rate constant of main power reactions for YMPQF oscillator, under control of R phase signal, is $104$. Simulation results are given in Fig. \ref{fig14}.

\qquad \textbf{\textit{Results and Analysis.}} As shown in Fig. \ref{fig14:a}, the red-colored final clock signal of $1/15$ duty cycle synthesized with Method 2 owns shorter time period than its blue-colored fundamental frequency. From Fig. \ref{fig14:b}, the $K$-phase oscillator, namely the $3$-phase one, produces a fundamental frequency of $1/3$ duty cycle. As an enlarged version of the red curve in Fig. \ref{fig14:a}, the bottom red curve in Fig. \ref{fig14:b} really segments one phase signal of $3$-phase oscillator into $5$ pieces. Thus the final clock signal colored red realizes the frequency multiplication of $1/15$ duty cycle.

\subsubsection{\textbf{Frequency Multiplication When $K=L$}}
Conditions of ``$K=L > 2$'' and ``$K=L=2$'' are taken into consideration as follows.

\paragraph{\textbf{\textit{For $K=L > 2$}}} Two methods are merged into one. We still employ the gear model in Fig. \ref{figm} and slow down the rate constant of main power reactions for the controlled $L$-phase oscillation.

\begin{figure}[htbp]
\centering
\subfigure[Fundamental frequency and frequency multiplication.]{
\label{figsp1:a}
\includegraphics[width=0.75\linewidth]{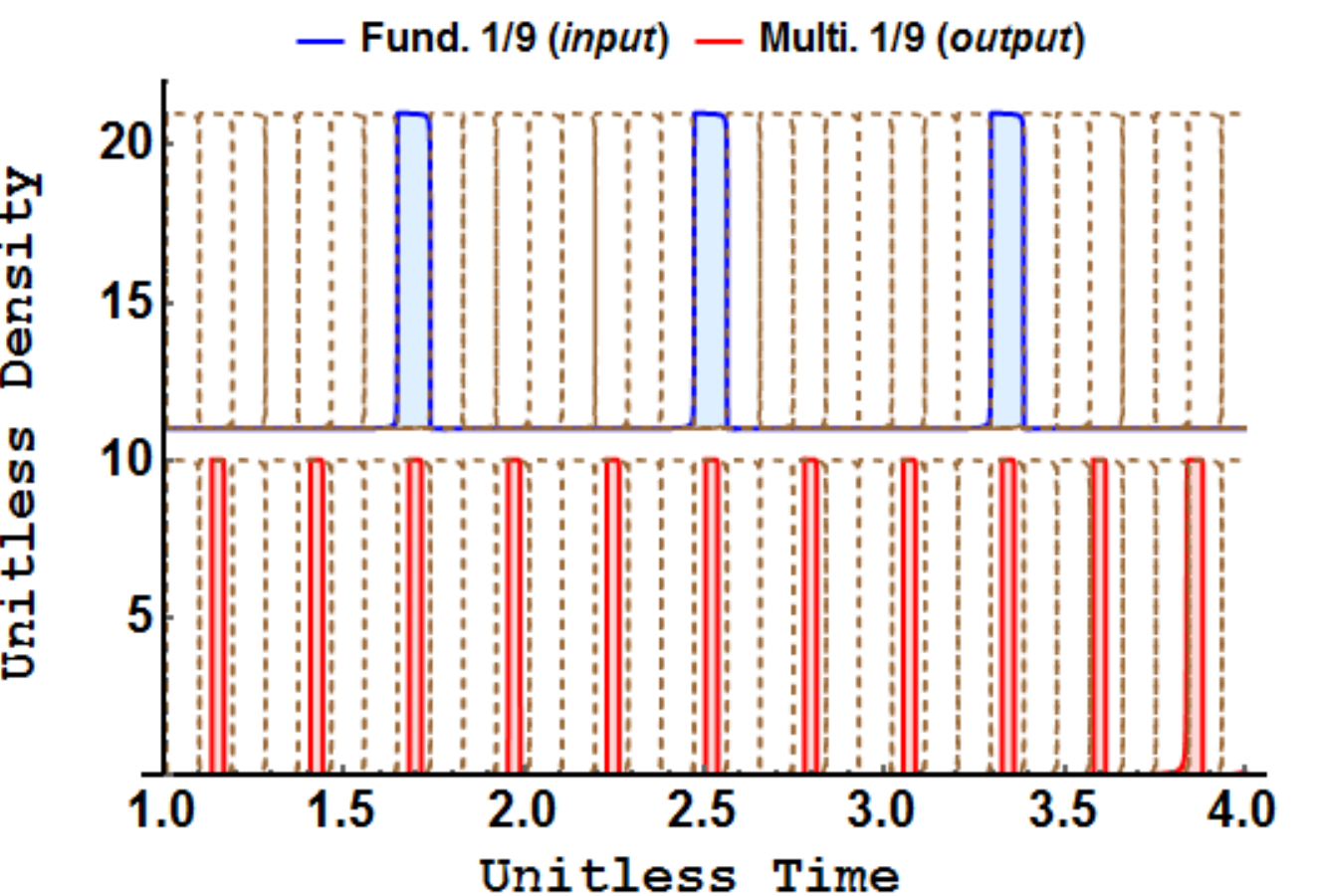}}
\subfigure[Output of $K$ oscillator and frequency multiplication.]{
\label{figsp1:b}
\includegraphics[width=0.75\linewidth]{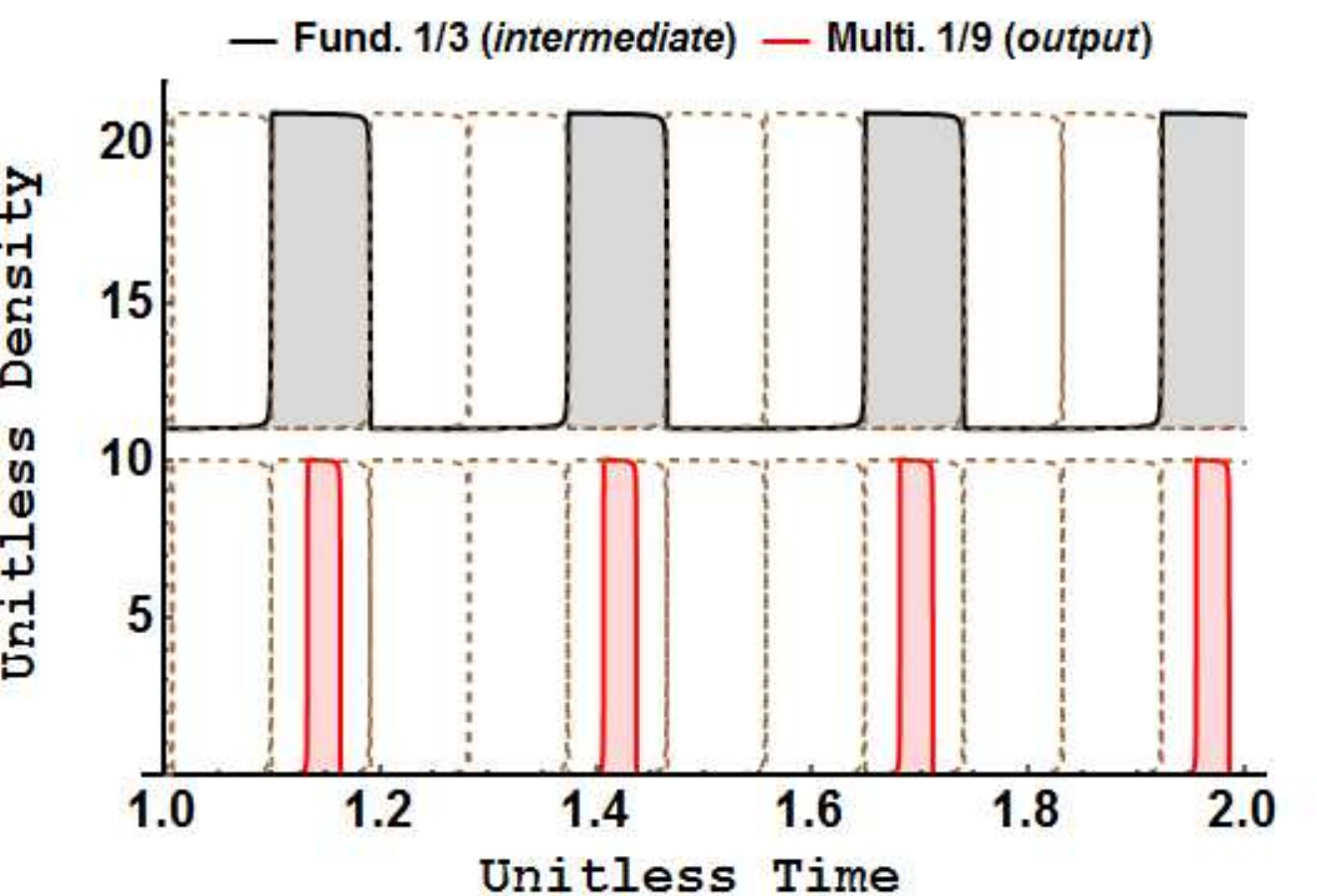}}
\caption{Simulation results for frequency multiplication when $K=L=3$.}\label{figsp1}
\end{figure}

\qquad \textbf{\textit{Example.}} Take a frequency multiplication of $1/9$ duty cycle as an example. The two oscillators are identical. The rate constant of main power reactions for $L$-phase oscillator is set to be $25.5$. Simulation results are shown in Fig. \ref{figsp1}. Finally, a nice frequency multiplication of $1/9$ duty cycle with only $36$ ($3 \!\! \times \!\! 4 \!\! \times \!\! 2 \!\! + \!\! 12 \!\! = \!\! 36$) reactions is well implemented in this way.

\begin{figure}[htbp]
\centering
\subfigure[Fundamental frequency and frequency multiplication.]{
\label{figk14:a}
\includegraphics[width=0.75\linewidth]{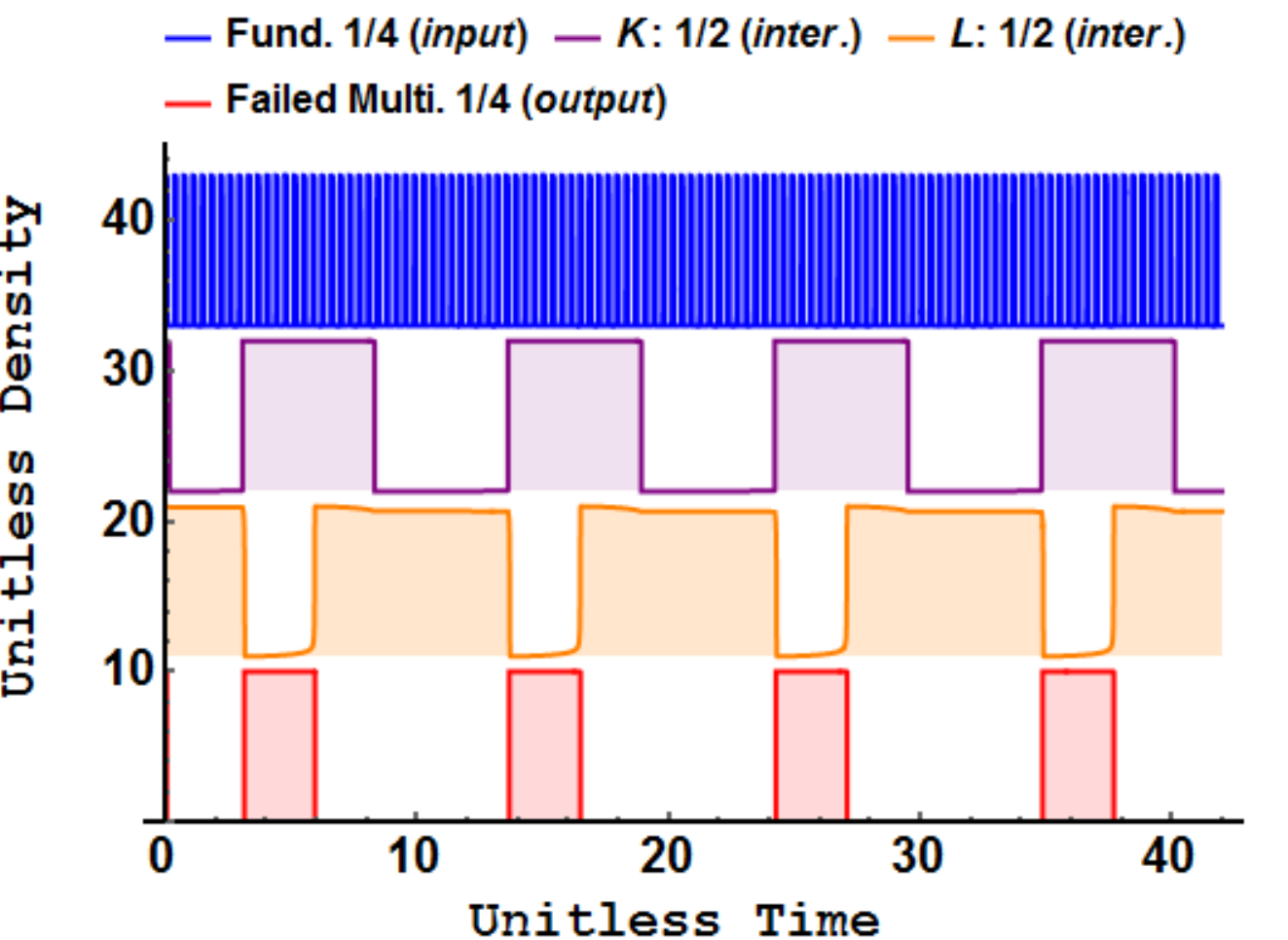}}
\subfigure[Output of $K$ oscillator and frequency multiplication.]{
\label{figk14:b}
\includegraphics[width=0.75\linewidth]{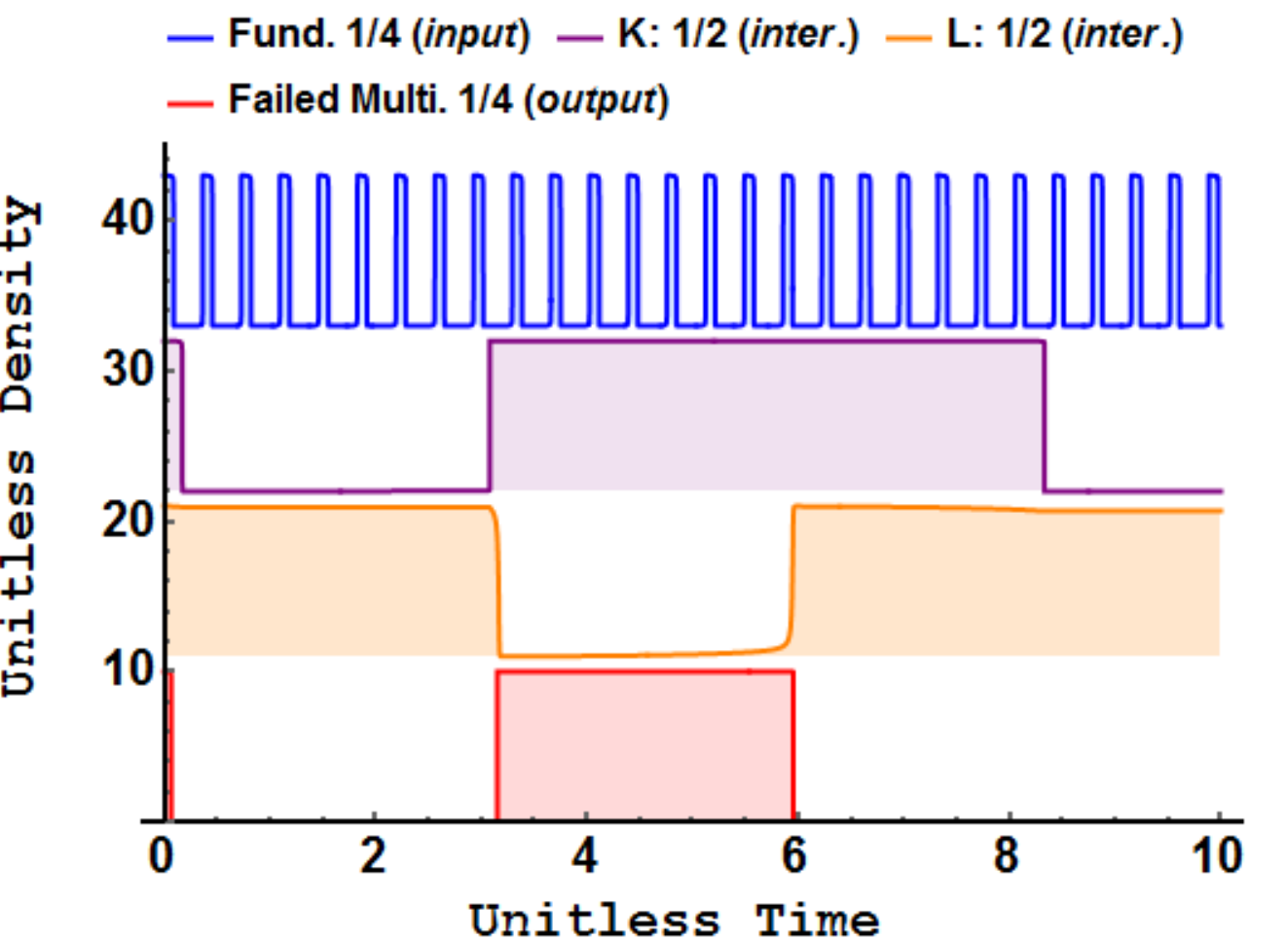}}
\caption{Simulation results for frequency multiplication when $K=L=2$.}\label{figk14}
\end{figure}

%\begin{figure}[htbp]
%\centering
%\subfigure[The output of gear system.]{
%\label{figsp2:a}
%\includegraphics[width=0.48\linewidth]{ca233.pdf}}
%\subfigure[Simulation of $ G_{2} $.]{
%\label{figsp2:b}
%\includegraphics[width=0.48\linewidth]{ca234.pdf}}
%\subfigure[The output of changed one.]{
%\label{figsp2:c}
%\includegraphics[width=0.48\linewidth]{ca231.pdf}}
%\subfigure[Simulation of the changed $ G_{2} $.]{
%\label{figsp2:d}
%\includegraphics[width=0.48\linewidth]{ca232.pdf}}
%\caption{Simulation results when $K$ or \(L = 2\). The top two figures show the simulation results without any change in chemical reactions. While simulation results when rate constant slowed down are shown at the bottom with the reduced value $2$. The phase signal of $ G_{2} $ are out of shape.}\label{figsp2}
%\end{figure}
\paragraph{\textbf{\textit{For $K=L=2$}}} Similar to the case of $K=L>2$, we should shorten the time period of $L$-phase oscillator. Although the rate constant scheme for this $L$-phase oscillator (threshold: $0.1$, main power: $0.4$) really shortens the time period, the proposed final $1/4$ duty cycle clock signal is not the wanted frequency multiplication, but an essential frequency division. Simulations shown in Fig. \ref{figk14} can confirm this. Therefore, our proposed method is inefficient for frequency multiplication in the case of $K=L=2$.
%\paragraph{\textbf{\textit{For $K$ or \(L = 2\)}}}

\subsubsection{\textbf{Frequency Multiplication When $K$ or \(L = 2\)}}
The proposed two methods lose their efficiency, but Method 1 could roughly still be in use. From Part I, the size of gear model for $1/2$ duty cycle is $37.8182$ times of that for $1/3$ duty cycle, or rather equals to that of standard $1/N$ duty cycle when $N=117$. In this sense, the standard fundamental frequency of $1/2$ duty cycle has a rather large time period than that of $L$ or $K$-phase oscillator. Therefore, the rate constant adjustment might not efficiently slow down or speed up the target oscillation.

\begin{figure}[htbp]
\centering
\subfigure[Fundamental frequencies.]{
\label{fig16d:a}
\includegraphics[width=0.48\linewidth]{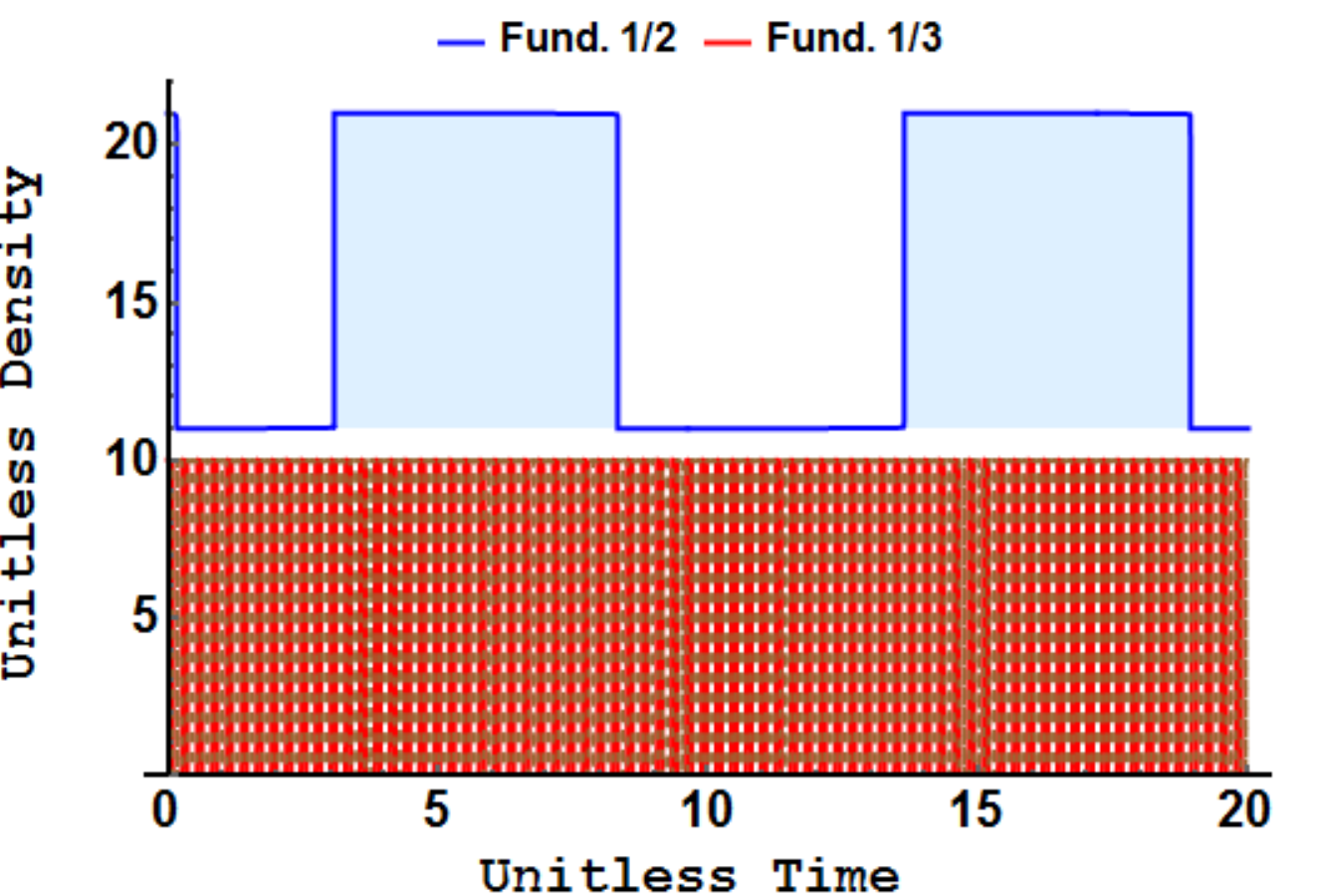}}
\subfigure[An enlarged version.]{
\label{fig16d:b}
\includegraphics[width=0.48\linewidth]{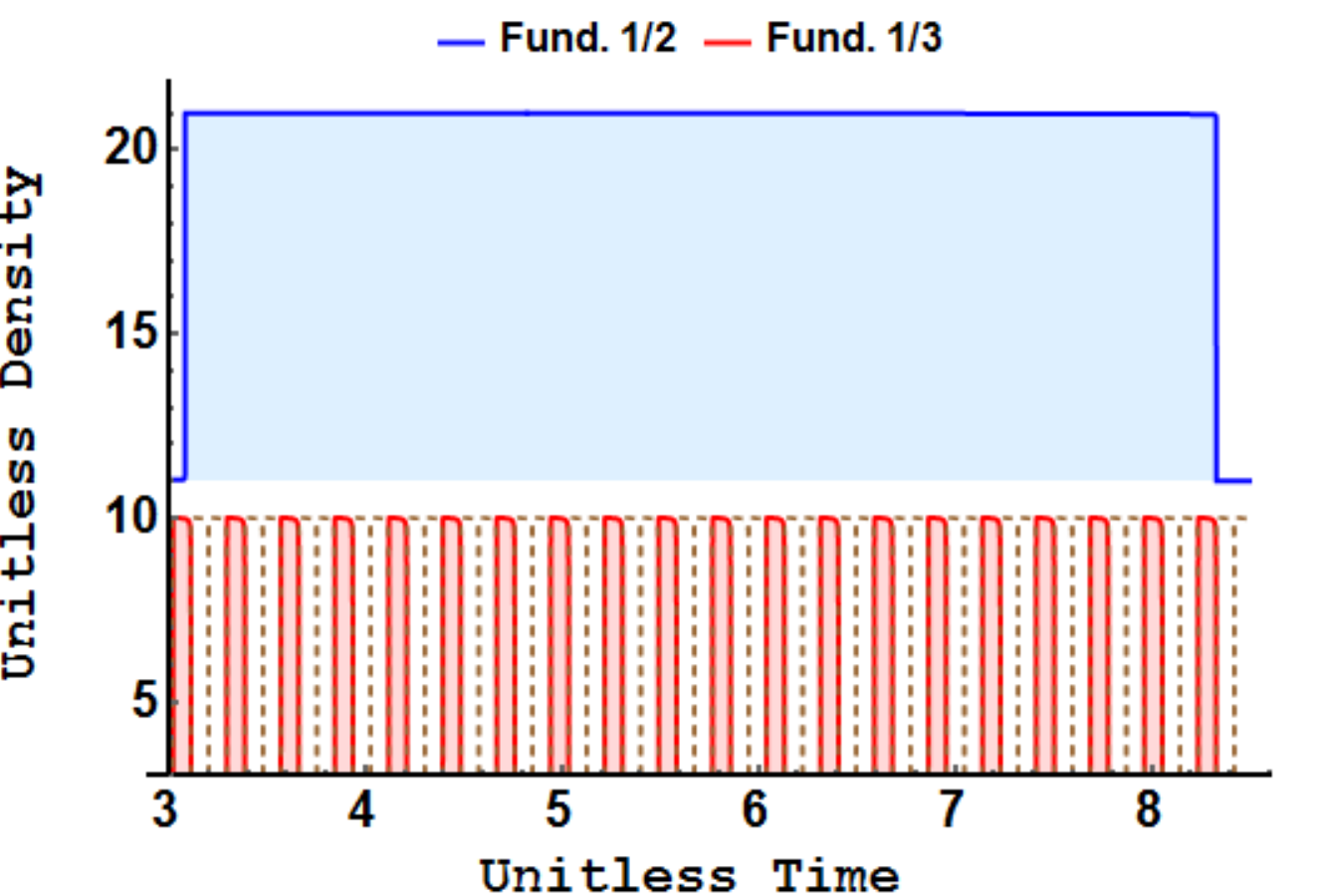}}
\caption{Results for fundamental frequencies for $1/2$ and $1/3$ duty cycle.}\label{fig16d}
\end{figure}

\quad \textbf{\textit{Example.}} Take a $1/6$ duty cycle for an example. The corresponding gear models are still employed those shown in Fig. \ref{figm}. Before conducting the frequency multiplication of $1/6$ duty cycle, fundamental frequencies for both $1/2$ and $1/3$ are given in Fig. \ref{fig16d}, from which the fundamental frequency of $1/2$ duty cycle has a bigger time period than that of $1/3$ one. This too fast transference of fundamental frequency for $1/3$ duty cycle makes CRNs for $1/3$ duty cycle have a longer time period in Method 1, while CRNs for $1/2$ duty cycle have a shorter one.

\begin{figure}[htbp]
\centering
\subfigure[Frequency multiplication for $1/6$ duty cycle (Meth. 1).]{
\label{fig16dm1:a}
\includegraphics[width=0.75\linewidth]{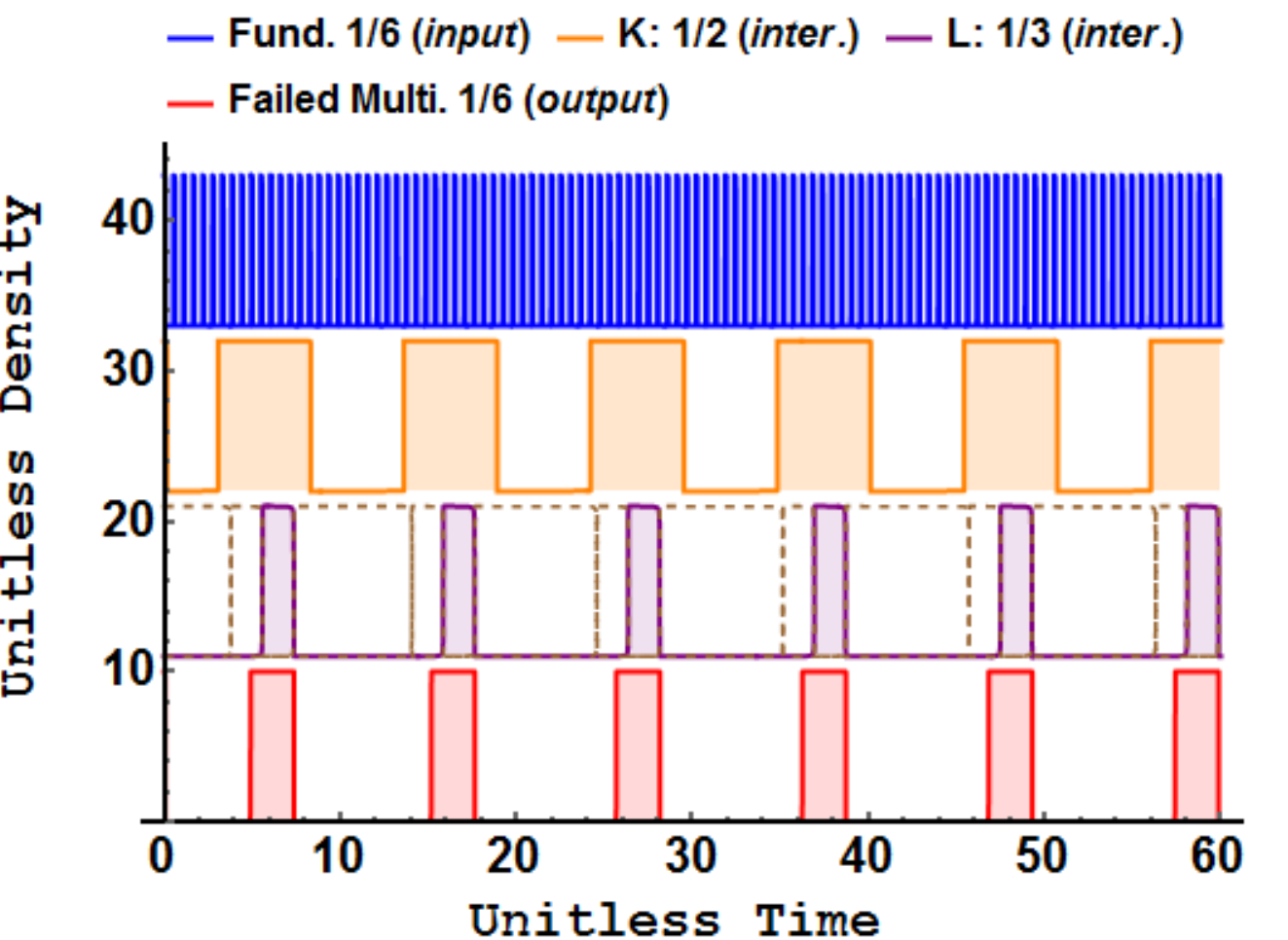}}
\subfigure[An enlarged version of the figure above.]{
\label{fig16dm1:b}
\includegraphics[width=0.75\linewidth]{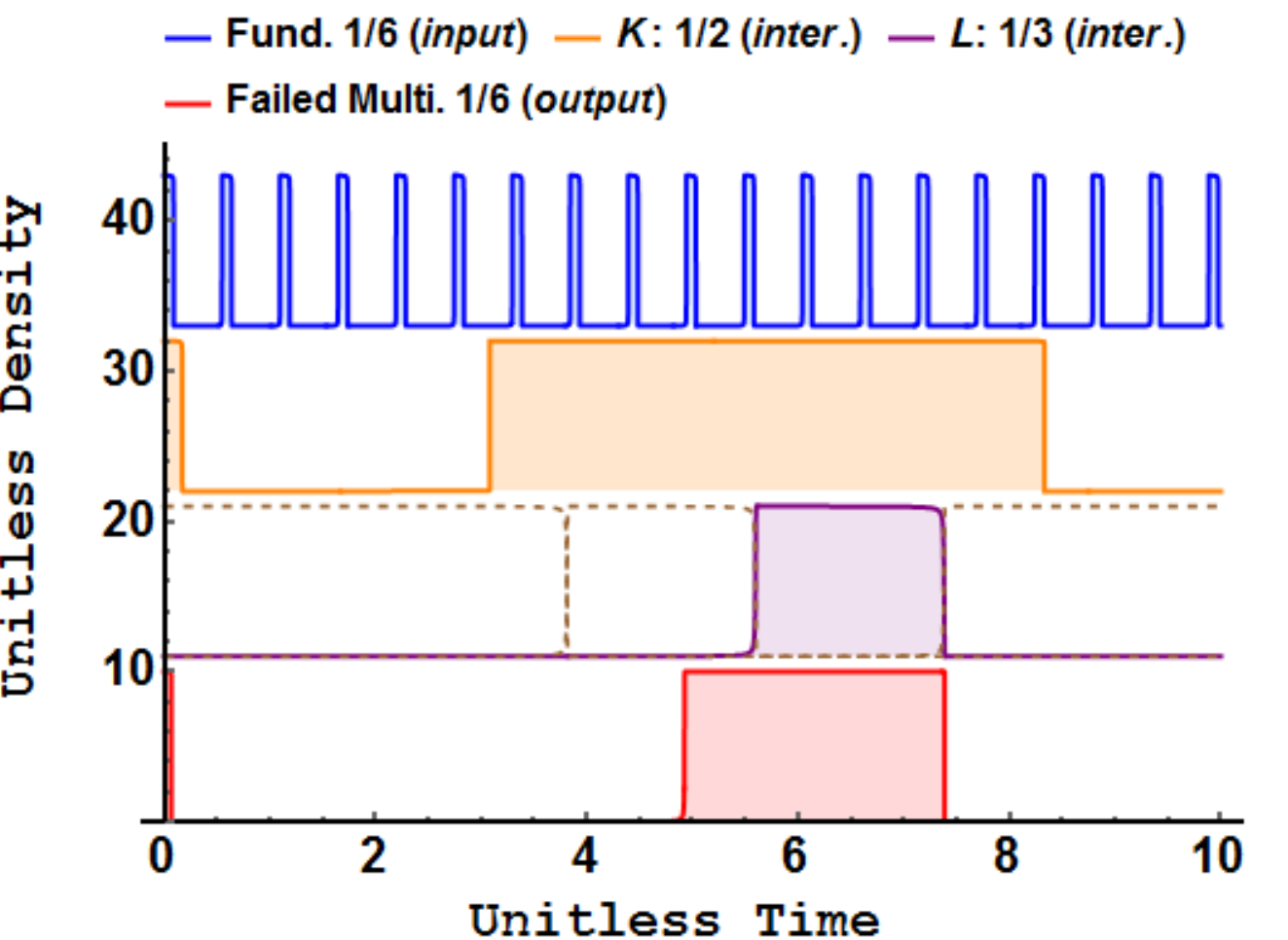}}
\caption{Simulations for frequency multiplication of $1/6$ duty cycle ($K=2$).}\label{fig16dm1}
\end{figure}

\paragraph{\textbf{\textit{For $K=2$}}} Adopted Method 1, $G_{A1}$ in Fig. \ref{figm} represents $1/2$ duty cycle. Employing the rate constant adjustment scheme that only rate constants for $1/3$ duty cycle oscillator are changed (threshold: $0.0002$, main power: $0.7$) while others still adopt the standard parameters in Part I, simulations are shown in Fig. \ref{fig16dm1}. From Fig. \ref{fig16dm1:a}, the final frequency multiplication of $1/6$ duty cycle seems to be good enough, but its enlarged version reveals that the segmentation of the second orange curve is not as precise as before. And the final red-colored frequency multiplication of $1/6$ duty cycle is a little bigger than the curve colored purple, although theoretically they should be the same size, as well as the same time period. To some extent, this error could be omitted. One thing should be emphasized is that, although this Method 1 seems to successfully construct a $1/6$ duty cycle clock signal, it is essentially a frequency division of the fundamental frequency, rather than frequency multiplication. This can be figured out when carefully comparing the ``wanted'' bottom red curve of Fig. \ref{fig16dm1} and the top blue curve of fundamental frequency. \textcolor{black}{Thus in this sense, we roughly think Method 1 for the case of $K=2$ is feasible for frequency division, but invalid for frequency multiplication.}

\begin{figure}[htbp]
\centering
\subfigure[Frequency multiplication for $1/6$ duty cycle (Meth. 2).]{
\label{fig16dm2:a}
\includegraphics[width=0.75\linewidth]{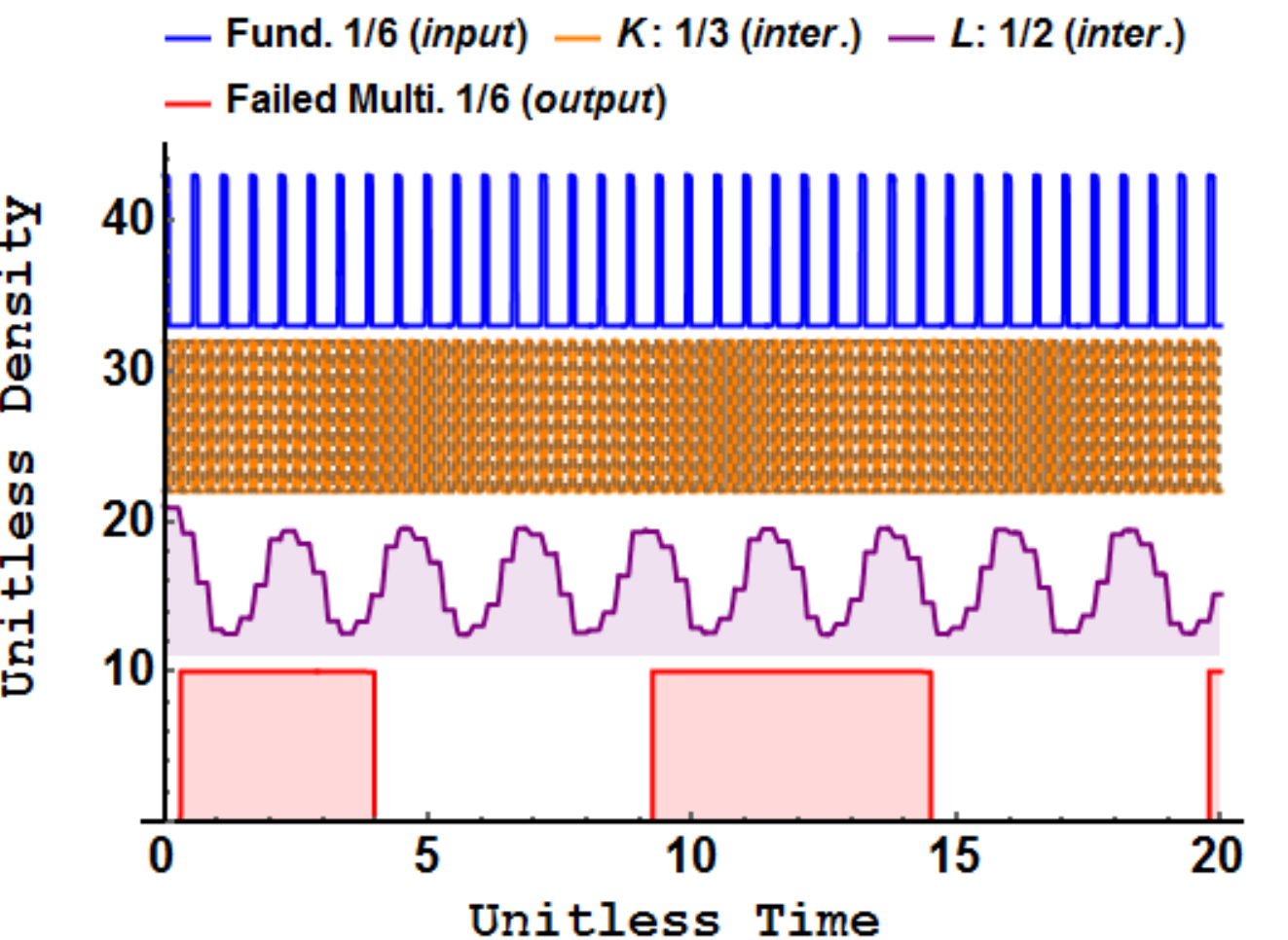}}
\subfigure[An enlarged version of the figure above.]{
\label{fig16dm2:b}
\includegraphics[width=0.75\linewidth]{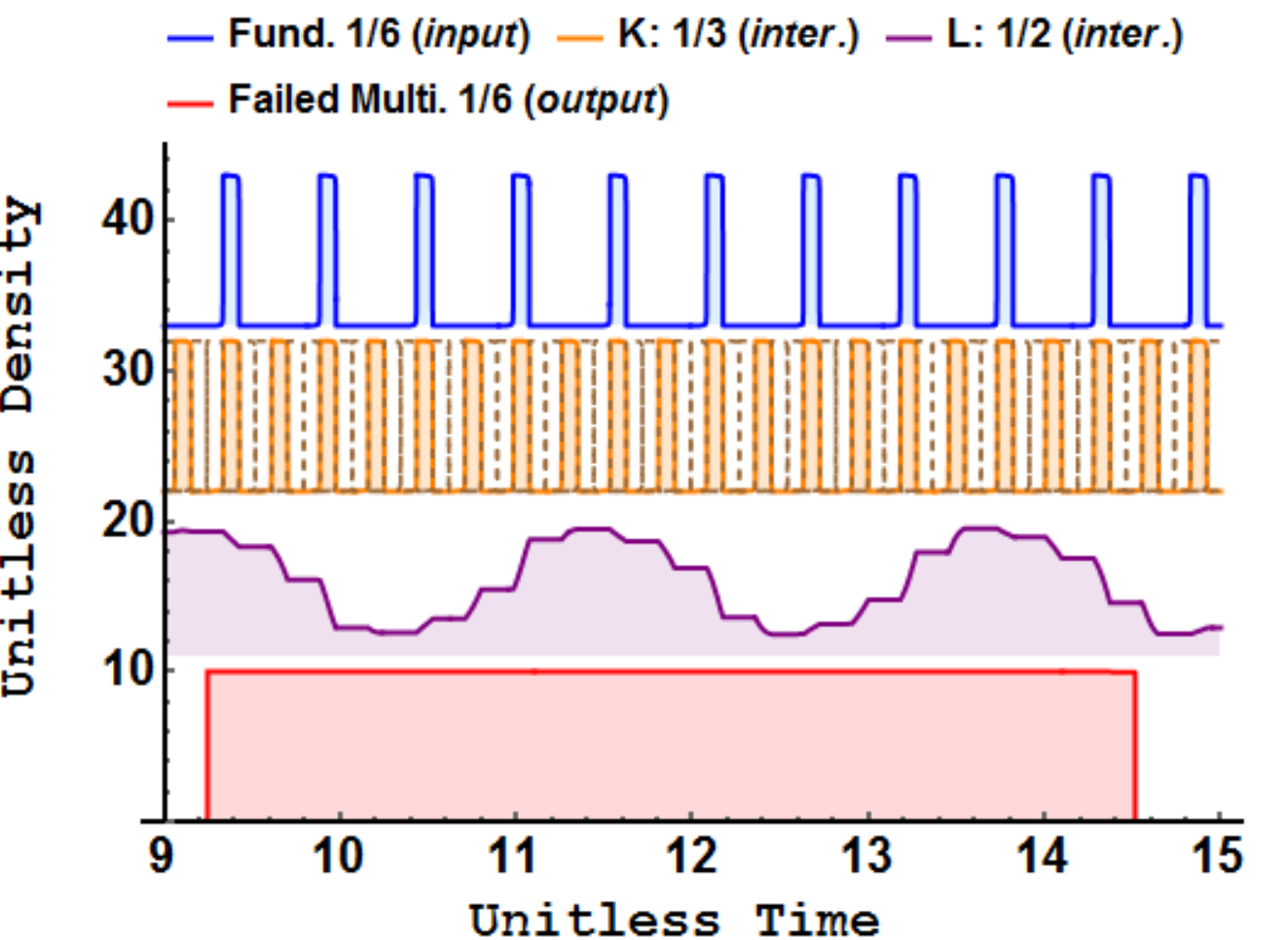}}
\caption{Simulations for frequency multiplication of $1/6$ duty cycle ($L=2$).}\label{fig16dm2}
\end{figure}

\paragraph{\textbf{\textit{For $L=2$}}} Adopted Method 2, then $G_{A2}$ in Fig. \ref{figm} represents $1/3$ duty cycle. To shorten the time period of $1/2$ duty cycle, the rate constant for main power reactions of $G_{A2}$ should be smaller than before. When it is set to be $0.08$, although the time period of $1/2$ duty cycle clock signal is really slowed down, it is not short enough to segment a single phase of a fundamental frequency for $1/3$ duty cycle. Moreover, this rate constant adjustment scheme produces a non-square wave of $1/2$ duty cycle clock signal. Therefore, revealed by Fig. \ref{fig16dm2}, the final frequency multiplication of $1/6$ duty cycle makes nonsense of Method 2.

%in the fist beginning of the blue-filled single phase

%If no change exists in CRNs' rate constants before coupling, simulation results are on the top of Fig. \ref{figk21}, from which these two oscillators obviously ``miss mesh''. Because of the physical truth that period of $1/2$ duty cycle is much longer than that of $1/3$ duty cycle. Or it is to say, under the control of a single phase signal $a_1$, the transference among R, G and B is quite faster than wanted. \textcolor{red}{Thus, this too fast transference of RGB oscillator could not only segment the existing time of $a_1$ into three pieces.} To this end, rate constant of CRNs for $1/3$ duty cycle should be slowed down. The corresponding simulation results are shown at the bottom of Fig. \ref{figk21}. From these results, we can find that, although the rate constant has been slowed down to $2$, the transference of RGB conducts more than once in a phase time of $a_1$. Therefore, this method does not work well.

%The rate constant of CRNs for $1/2$ duty cycle with two phases of $a_1$ and $a_0$ should be increased. Simulation results are shown in Fig. \ref{figk22}. Although the transference of $1/2$ duty cycle has been faster and costs less time, the whole transference is not short enough to segment a phase time of RGB oscillator. Therefore, this method still does not work well. In short, in condition of $K$ or \(L = 2\), our proposed methods could not work well.

%\begin{figure}[htbp]
%\centering
%\subfigure[The output of gear system.]{
%\label{figk21:a}
%\includegraphics[width=0.48\linewidth]{dkxa3.pdf}}
%\subfigure[The output of gear system.]{
%\label{figk21:b}
%\includegraphics[width=0.48\linewidth]{dkxa4.pdf}}
%\subfigure[The output of gear system.]{
%\label{figk21:c}
%\includegraphics[width=0.48\linewidth]{dkxa1.pdf}}
%\subfigure[Results of RGB and $R_{0}R_{1}R_{2}$.]{
%\label{figk21:d}
%\includegraphics[width=0.48\linewidth]{dkxa2.pdf}}
%\caption{Simulation results when $K=L=3$.}\label{figk21}
%\end{figure}

%\begin{figure}[htbp]
%\centering
%\subfigure[The output of gear system.]{
%\label{figk22:a}
%\includegraphics[width=0.48\linewidth]{xkdb1.pdf}}
%\subfigure[Results of RGB and $R_{0}R_{1}R_{2}$.]{
%\label{figk22:b}
%\includegraphics[width=0.48\linewidth]{xkdb2.pdf}}
%\caption{Simulation results when $K$ or \(L = 2\).}\label{figk22}
%\end{figure}

\subsubsection{\textbf{Summarization}}
In our proposal, based on the compound gear model, the frequency multiplication could be successfully implemented. Additionally, an appropriate rate constant adjustment should be adopted for a better final result. The basic idea to realize frequency multiplication of $1/N$ is illustrated in Algorithm \ref{arg2}.

\begin{algorithm}
\caption{Methods for frequency multiplication of $1/N$.}\label{arg2}
\begin{algorithmic}[1]
\REQUIRE A compound gear model of three oscillators.
\IF {$N=K \times L$ is a composite number (\(K,L \ne 1\))}
\STATE The three oscillators are $1/2$ , $1/K$ and $1/L$ duty cycle.
\ENDIF
\IF {\(K \neq L > 2\)}
\STATE Use a phase signal of $K$-phase oscillator to control the whole transference of $L$-phase one.
\STATE Two phase signals of $L$-phase oscillator are used to control $1/2$ duty cycle.
\STATE (\textit{\textbf{Method 1.}}) For $K>L$, slow down the rate constant of main power reactions of the controlled $L$-phase oscillator.
\STATE (\textit{\textbf{Method 2.}}) For $K<L$, speed up the rate constant of main power reactions of controlled $L$-phase oscillator.
\ELSIF {$K=L$}
\STATE Two methods are merged into a single method.
\IF {$K=L>3$}
\STATE This situation is similar to Method 1.
\ELSIF {$K=L=2$}
\STATE Our methods are inefficient.
\ENDIF
\ELSIF {$K$ or \(L = 2\)}
\STATE \textcolor{black}{Invalid methods}.
\ENDIF
%\IF{For the sake of frequency multiplier}
%\STATE Frequency multiplier could be well achieved through speeding up the rate constant of chemical reactions.
%\ENDIF
\end{algorithmic}
\end{algorithm}

Note that, the mentioned methods in Algorithm \ref{arg2} could not only implement a $1/N$ duty cycle clock signal with fewer reactions, but also construct an $M/N$ one, with a restriction that $M$ must be less (greater) than $L/N$ ($1-L/N$) in Method 1 or less (greater) than $K/N$ ($1-K/N$) in Method 2. This is because we use one phase of an oscillator to control the rotation of another one, only $1/K$ ($1/L$) time period of the former clock signal could be segmented into $L$ ($K$) pieces, and duty cycle of $1/N$, $2/N$,..., $1/L$ ($1/K$) could be realized. The corresponding dual clock signals, namely duty cycle of $(N-L)/N$ ($(N-K)/N$),..., $(N-1)/N$ could also be implemented.

\begin{figure}[htbp]
\vspace{-38pt}
\centering
\subfigure[Result of $1/15$ with Meth. 1.]{
\label{fig8:a}
\includegraphics[width=0.48\linewidth]{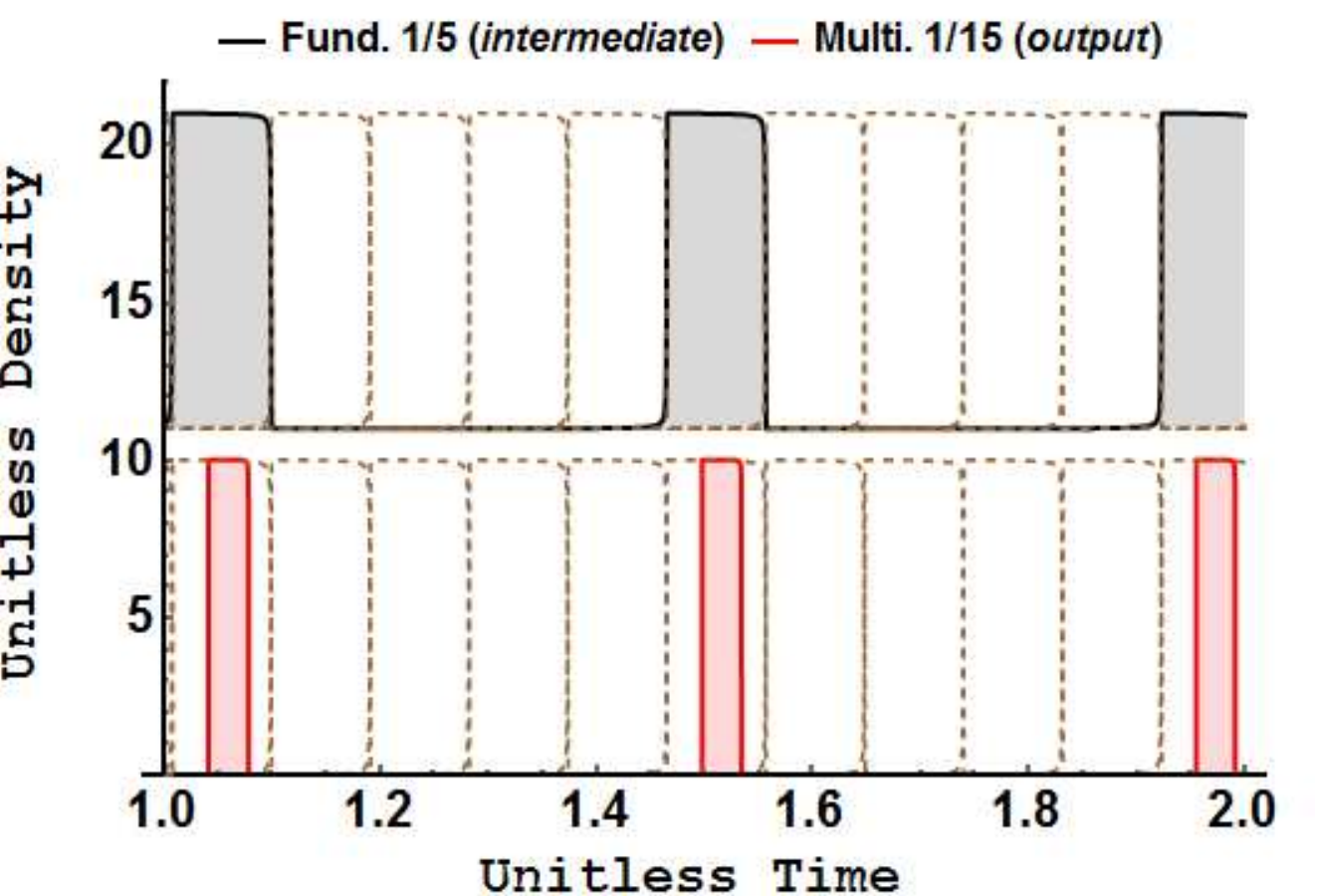}}
\subfigure[Result of $2/15$ with Meth. 1.]{
\label{fig8:b}
\includegraphics[width=0.48\linewidth]{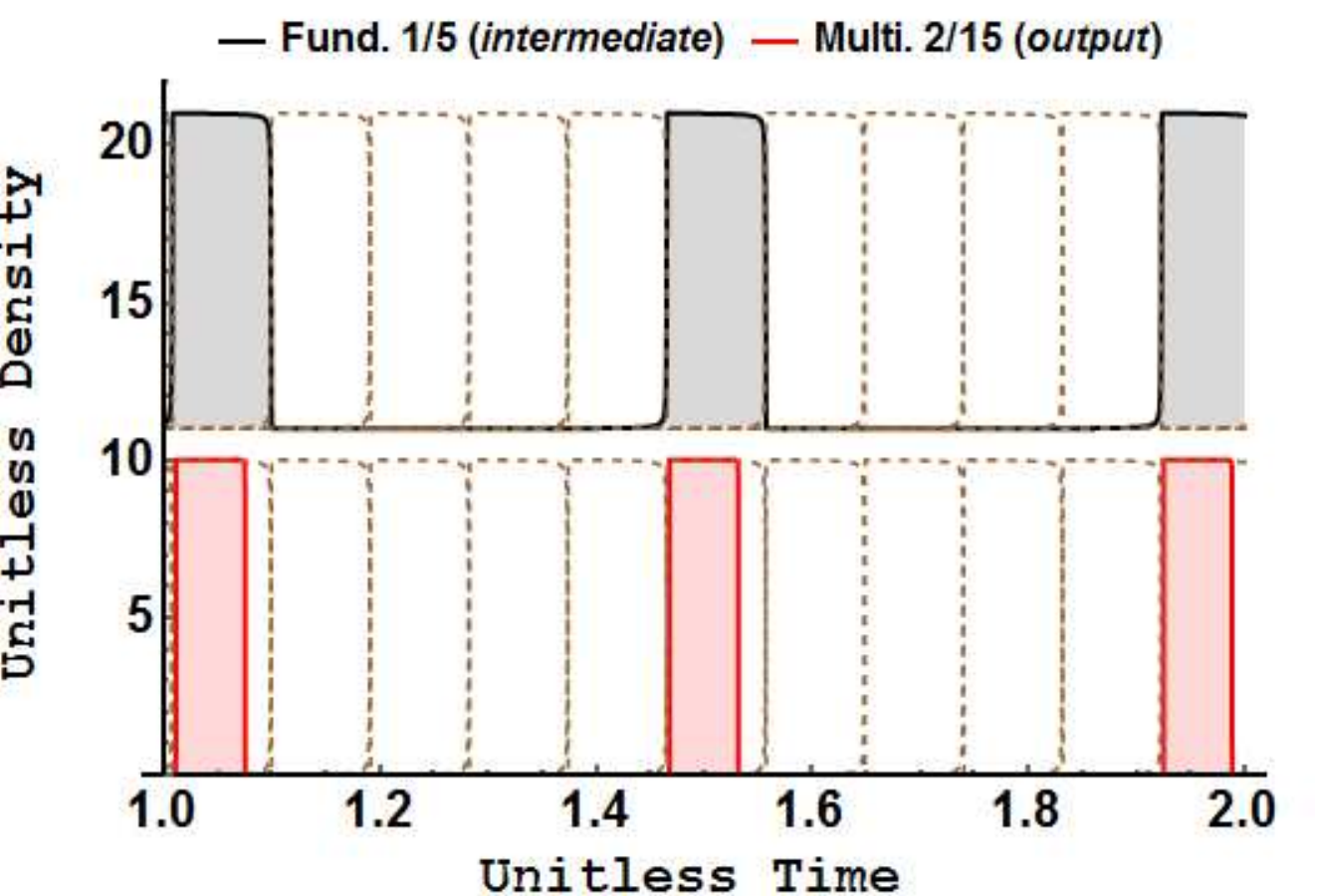}}
\subfigure[Result of $14/15$ with Meth. 1.]{
\label{fig8:c}
\includegraphics[width=0.48\linewidth]{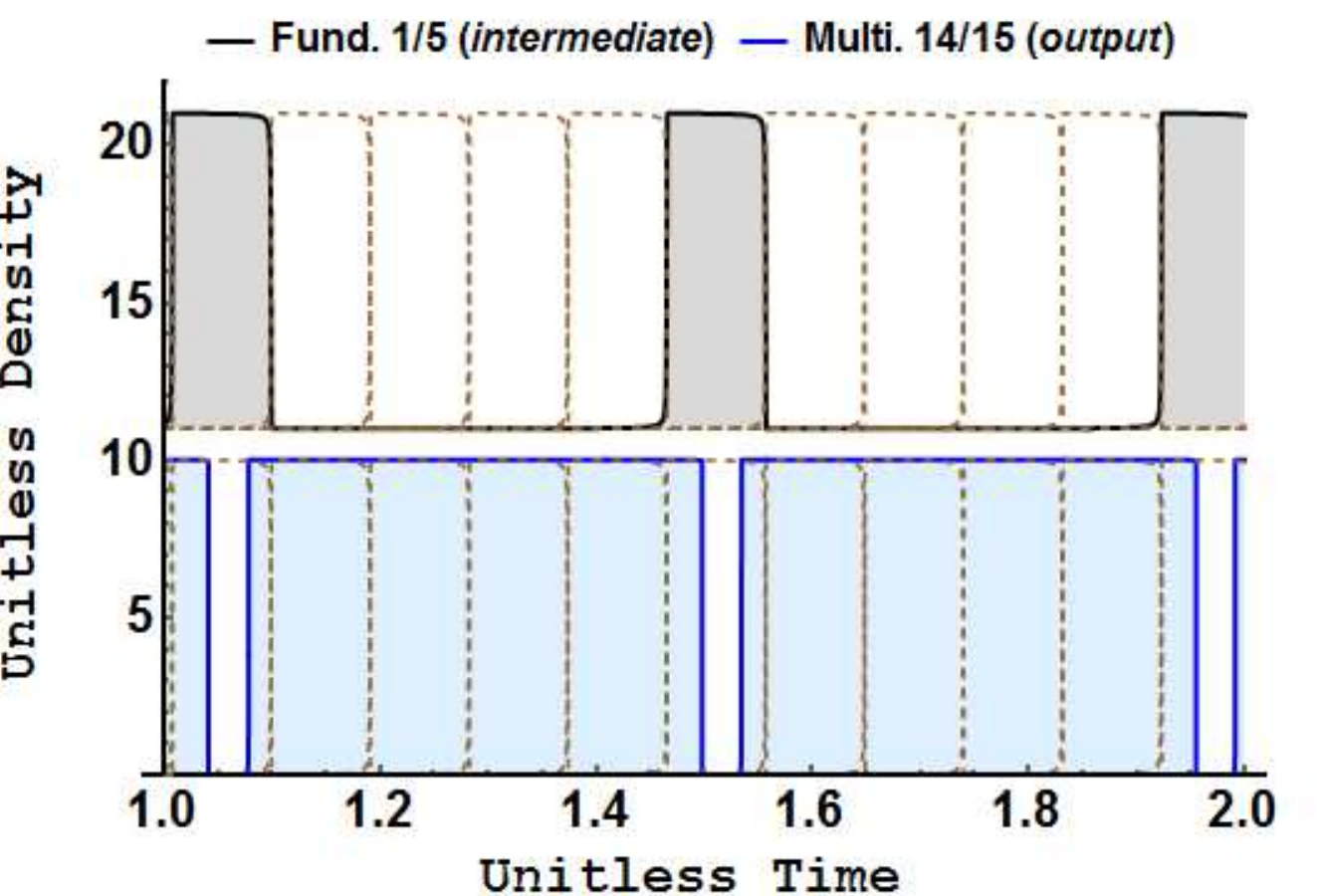}}
\subfigure[Result of $13/15$ with Meth. 1.]{
\label{fig8:d}
\includegraphics[width=0.48\linewidth]{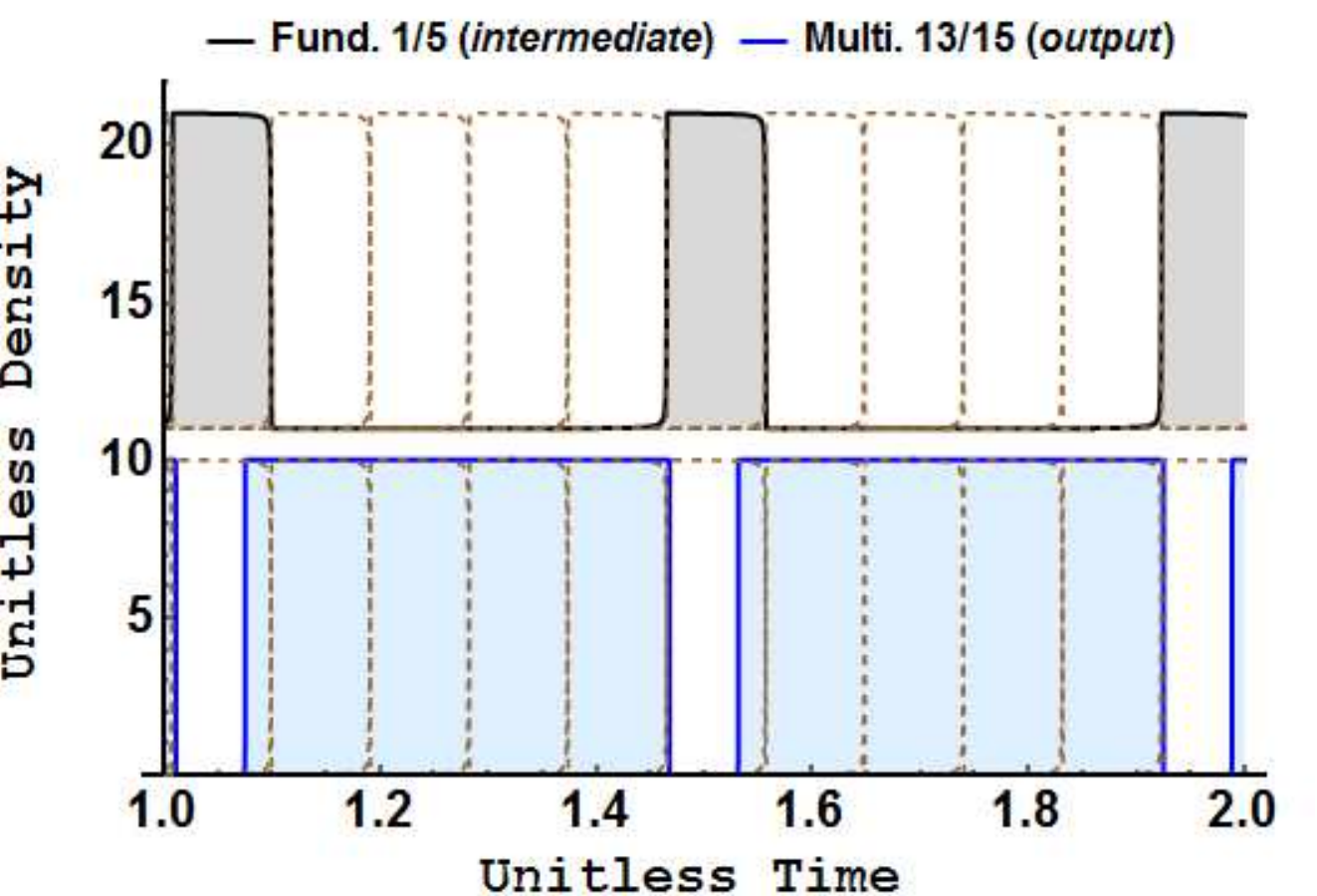}}
\subfigure[Result of $1/15$ with Meth. 2.]{
\label{fig8:e}
\includegraphics[width=0.48\linewidth]{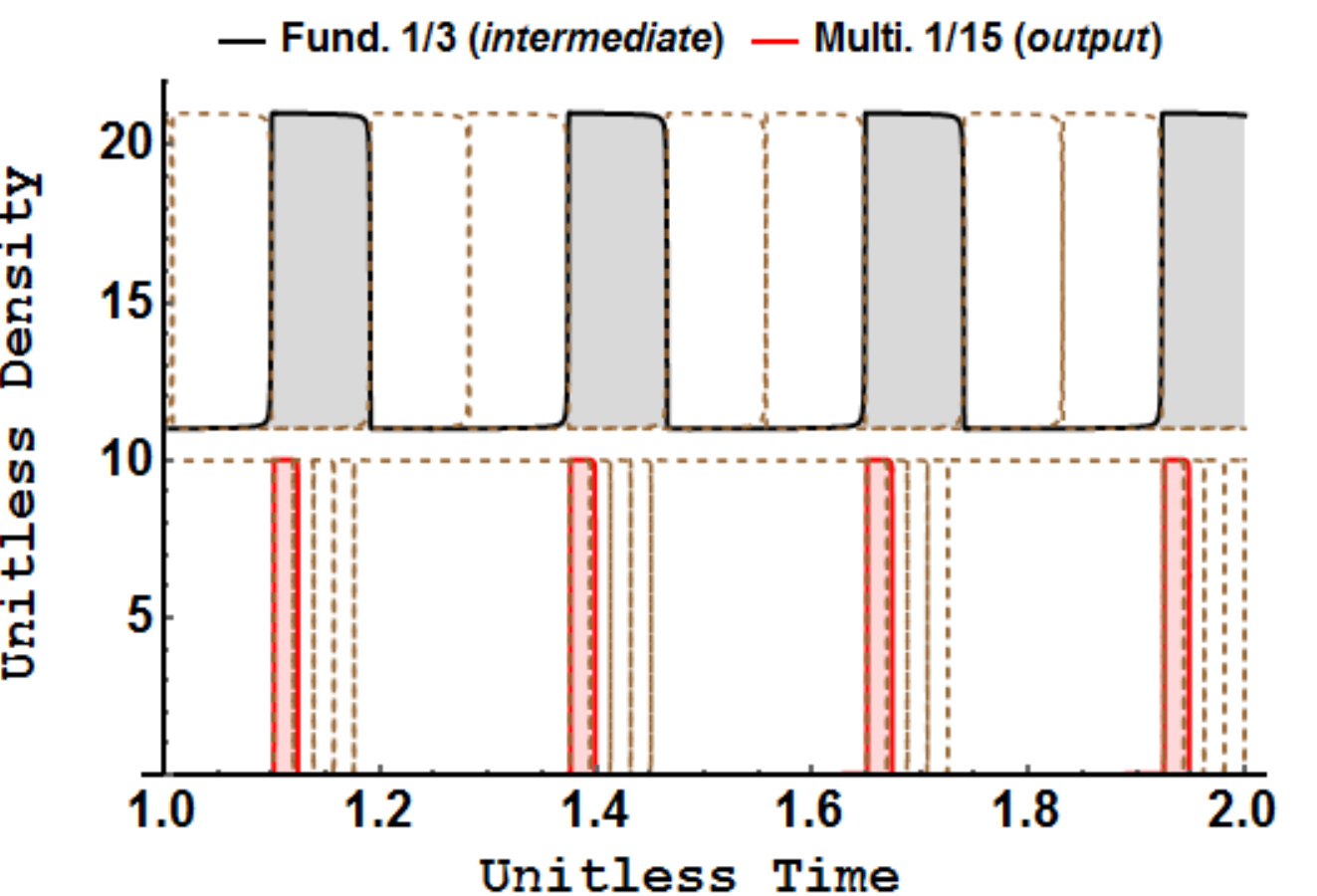}}
\subfigure[Result of $2/15$ with Meth. 2.]{
\label{fig8:f}
\includegraphics[width=0.48\linewidth]{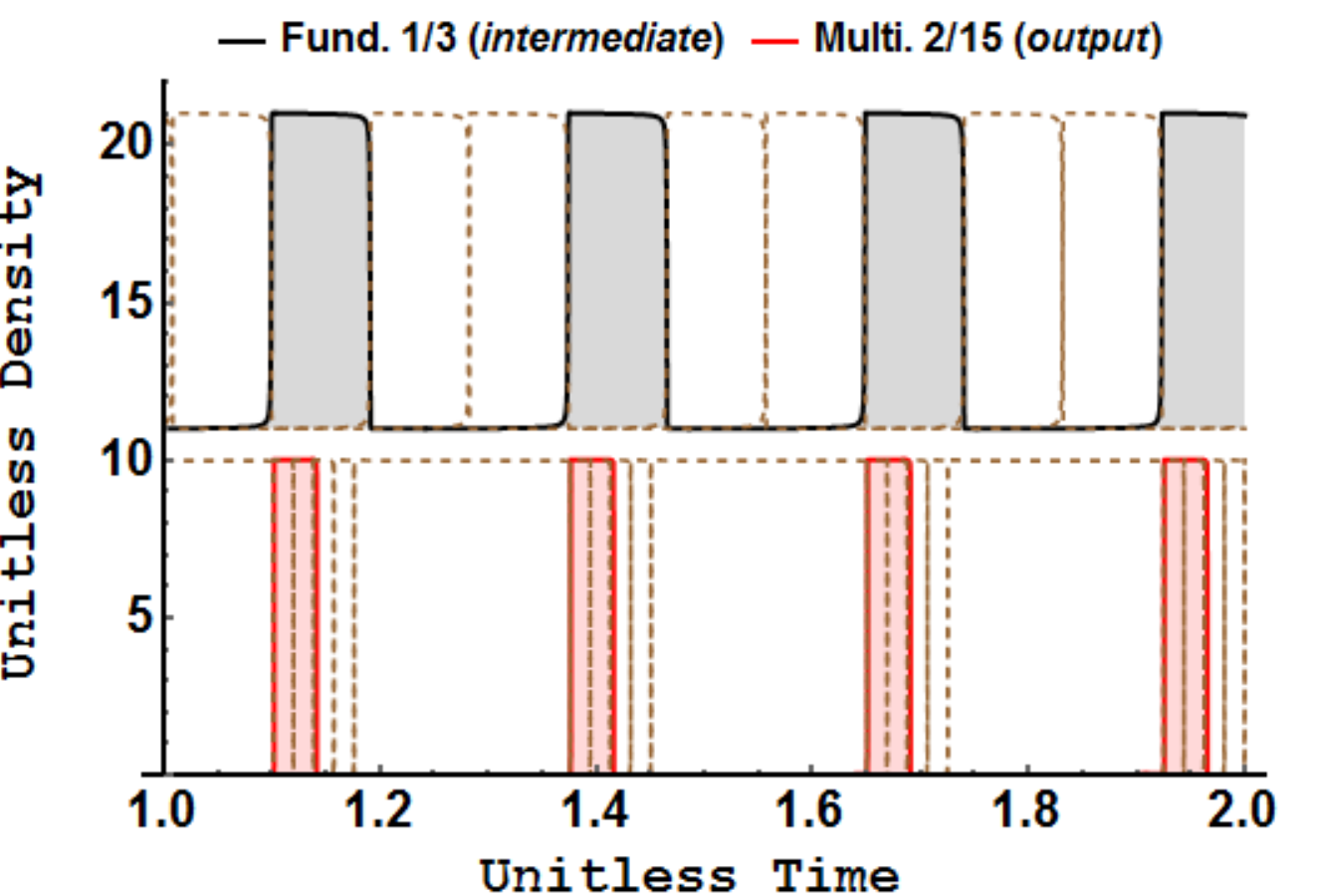}}
\subfigure[Result of $3/15$ with Meth. 2.]{
\label{fig8:g}
\includegraphics[width=0.48\linewidth]{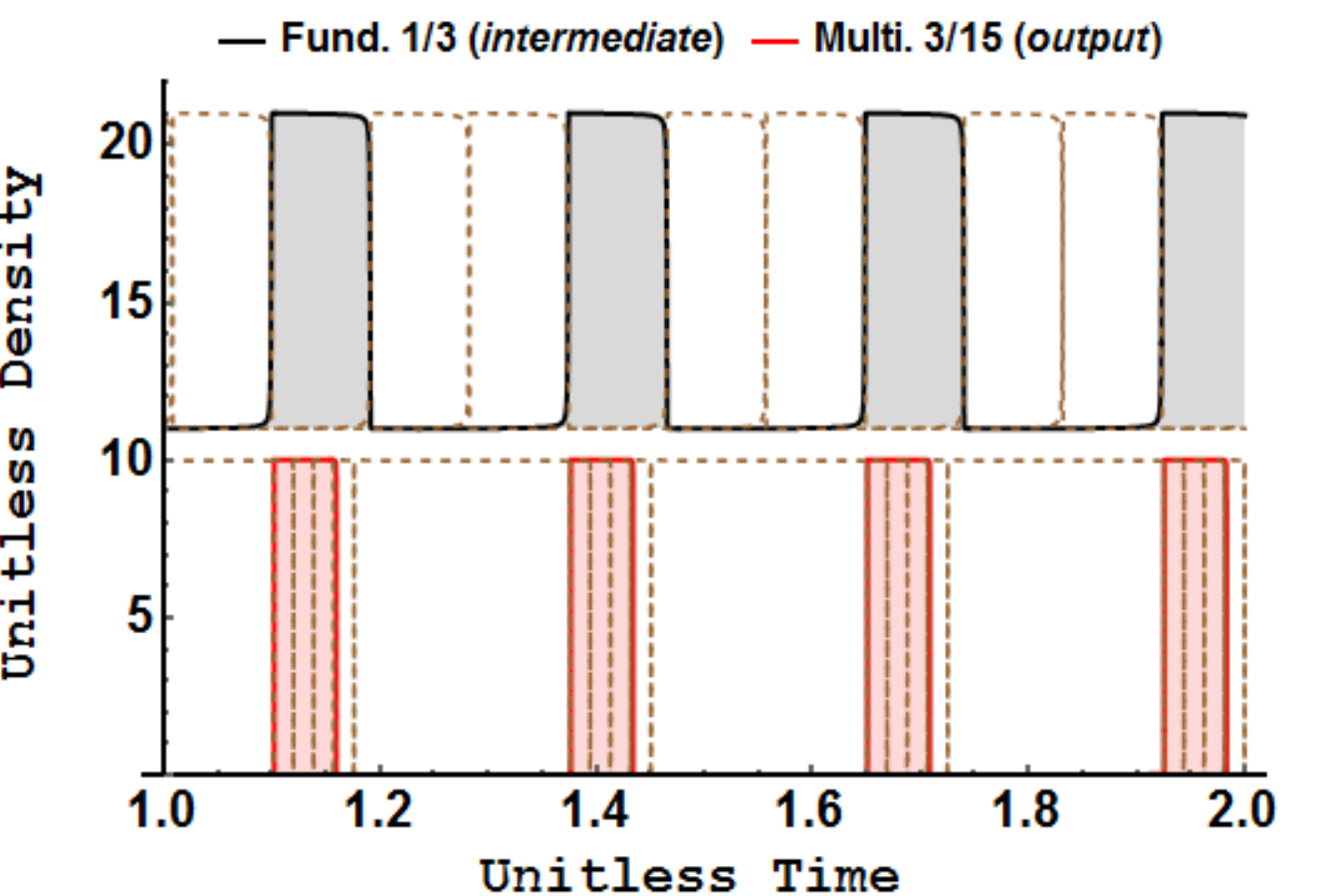}}
\subfigure[Result of $4/15$ with Meth. 2.]{
\label{fig8:h}
\includegraphics[width=0.48\linewidth]{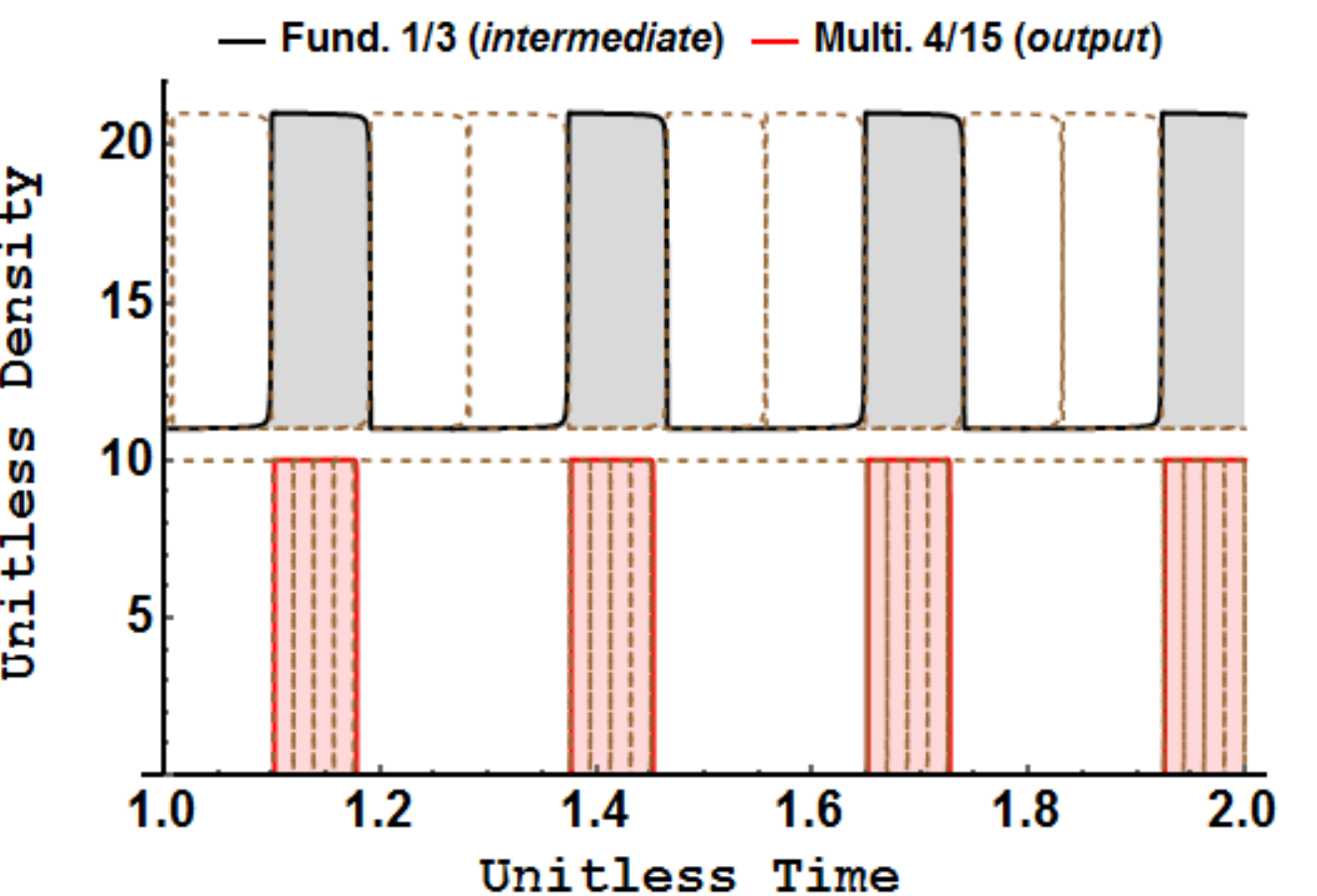}}
\subfigure[Result of $14/15$ with Meth. 2.]{
\label{fig8:i}
\includegraphics[width=0.48\linewidth]{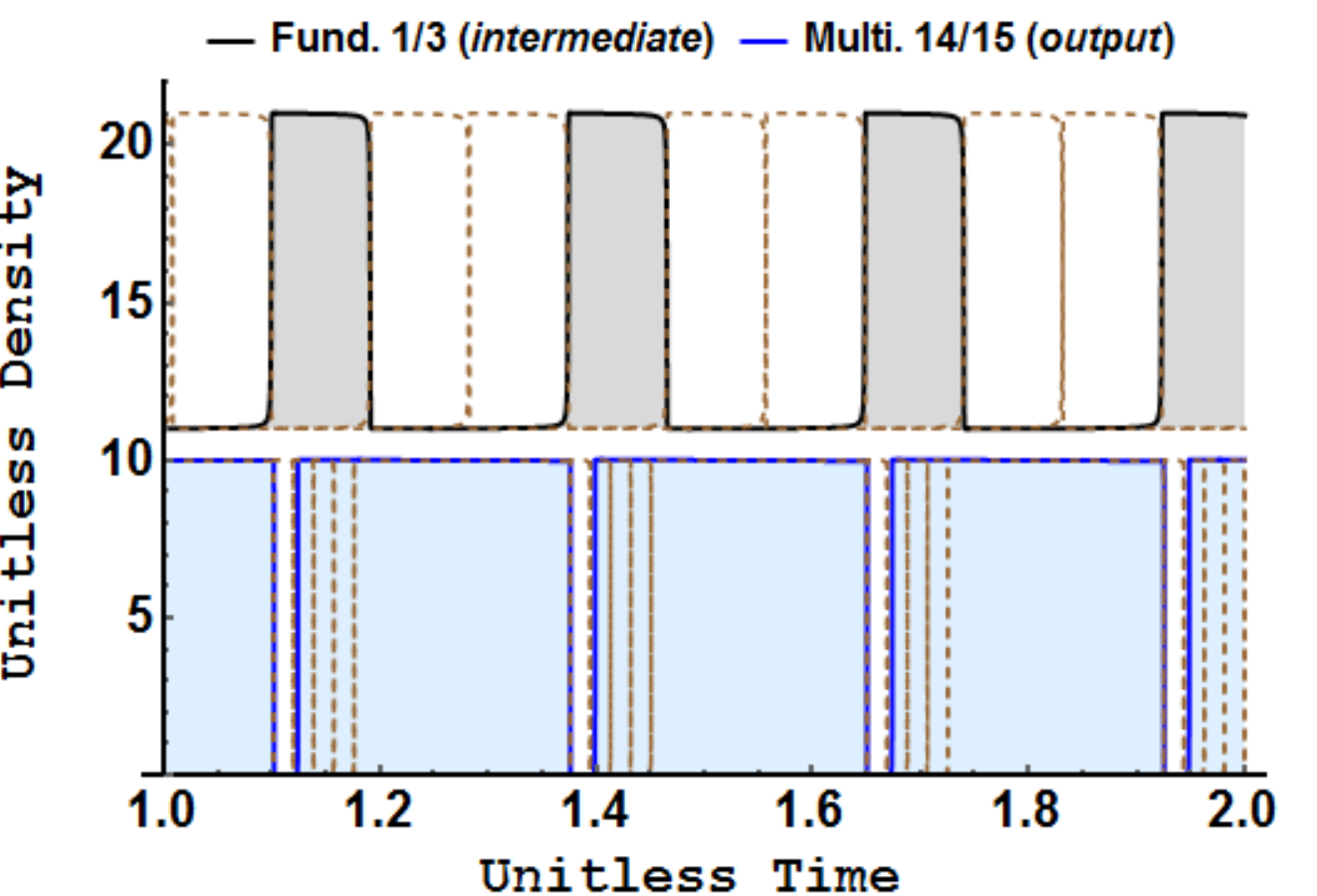}}
\subfigure[Result of $13/15$ with Meth. 2.]{
\label{fig8:j}
\includegraphics[width=0.48\linewidth]{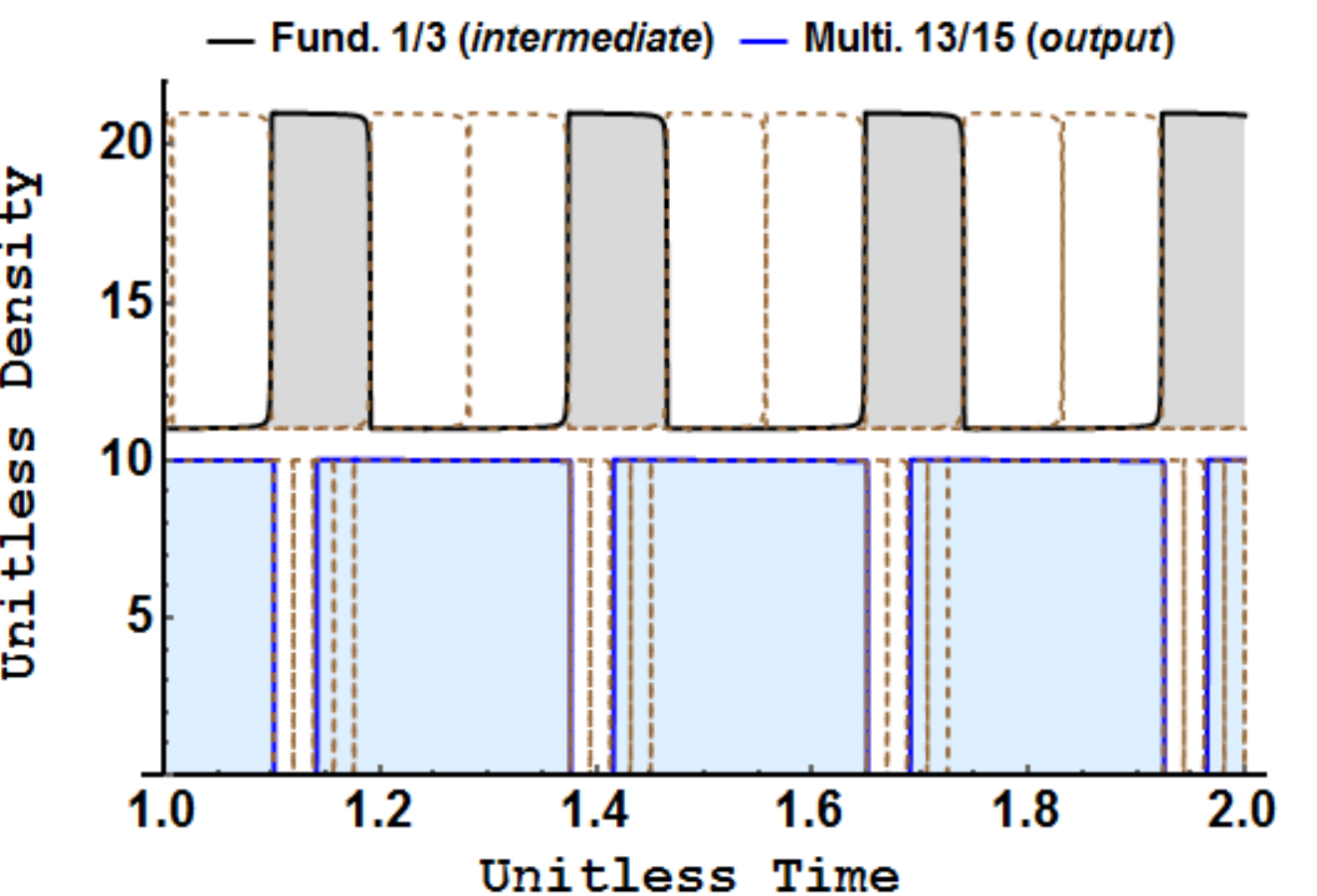}}
\subfigure[Result of $12/15$ with Meth. 2.]{
\label{fig8:k}
\includegraphics[width=0.48\linewidth]{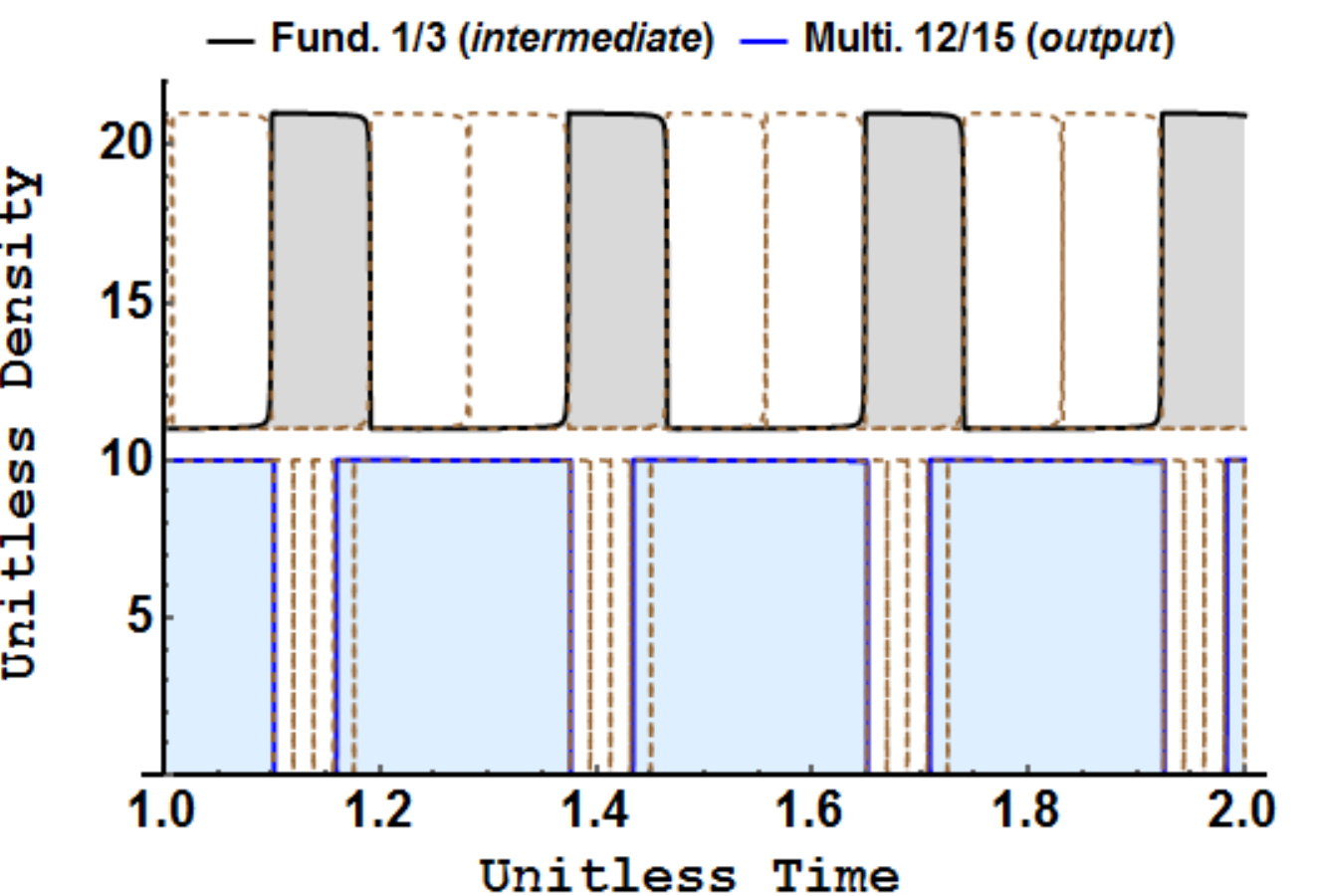}}
\subfigure[Result of $11/15$ with Meth. 2.]{
\label{fig8:l}
\includegraphics[width=0.48\linewidth]{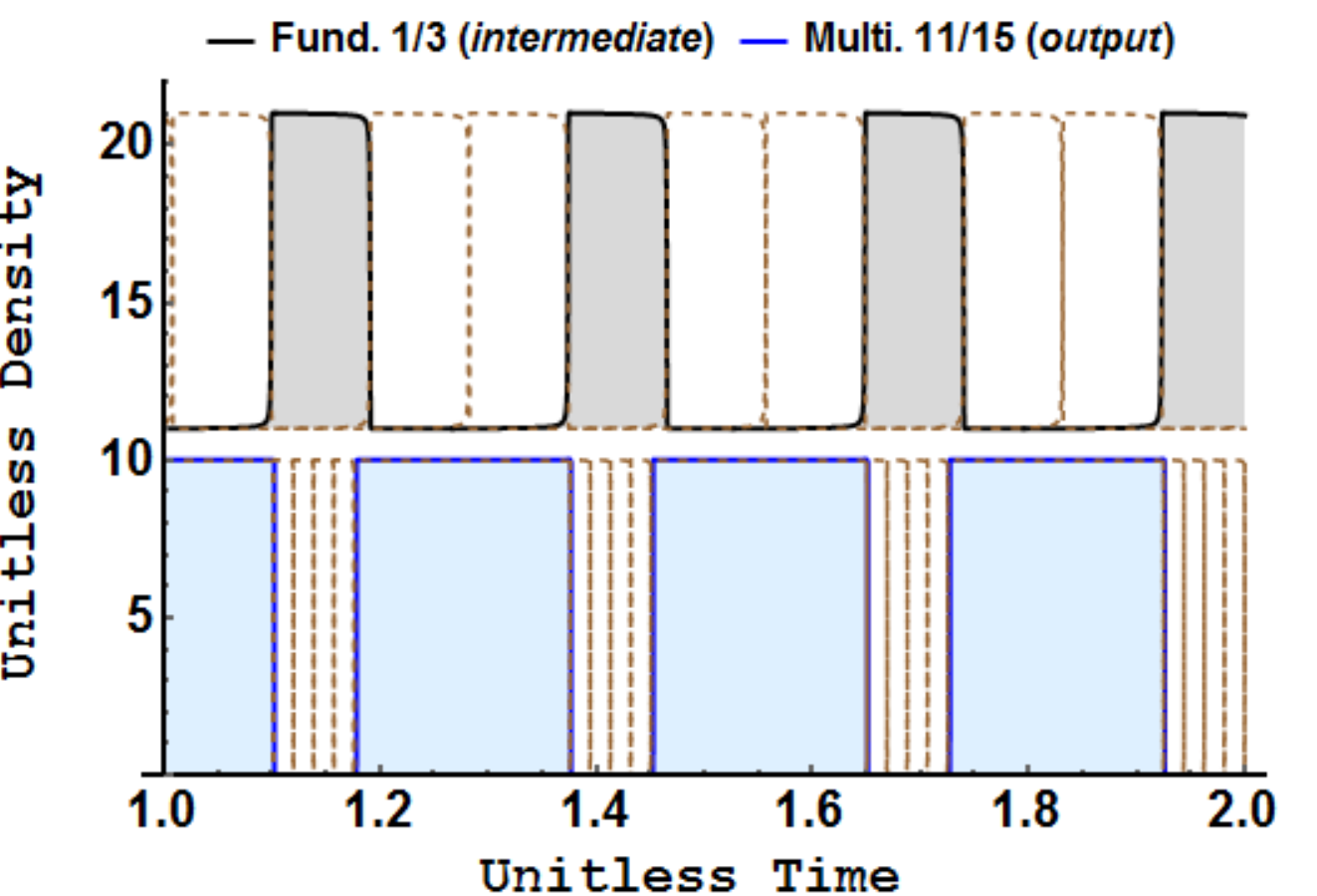}}
\caption{Simulation results of $1/15$, $2/15$ duty cycle with Method 1 or $1/15$, $2/15$, $3/15$, $4/15$ with Method 2, as well as their respective dual signals.}\label{fig8}
\vspace{-18pt}
\end{figure}

Take a $M/15$ duty cycle as an example, the other simulation results of $M/15$ duty cycle clock signals are shown in Fig. \ref{fig8}. That means, the model as shown in Fig. \ref{figm} could only be utilized to realize $1/15, 2/15$ duty cycle clock signal with Method 1, and their dual ones, namely $14/15, 13/15$. Moreover, simulation results for $1/15, 2/15, 3/15, 4/15$ as well as their dual ones of $14/15, 13/15, 12/15, 11/15$ with Method 2 are also shown at the bottom of Fig. \ref{fig8}. One thing should be emphasized is that the corresponding CRNs only requires \((4 \times (3+5)+12)=44\) chemical reactions. All of these prove that our gear model is meaningful to instruct us in further study and future applications of a clock tree in CRN level.

\section{Discussion and Conclusion}\label{sec:6}
In this paper, by exploiting the compound gear model in Part I, we could realize the frequency alteration by $L/J$, with frequency division and frequency multiplication. Another benefit is, with the accumulated instructions, we successfully use fewer chemical reactions to implement a clock signal with $M/N$ duty cycle, where $N$ may be very large. One thing should be emphasized is that, here $M$ does not range from $1$ to $N$, but has different ranges for Method 1 and Method 2. More conditions are also taken into considerations in our proposal, which has been summarized in Algorithm \ref{arg2}. With a bit slower or faster fine-tunings of rate constant, frequency multiplication could be well implemented based on our gear model, with fewer chemical reactions. In our previous work of \cite{ge2015formal}, nearly no quantitative description of the results was offered, however. Our simple model semi-quantitatively reproduces all the simulation data with a set of physically reasonable parameters.

\begin{figure}[htbp]
\centering
\includegraphics[width=0.8\linewidth]{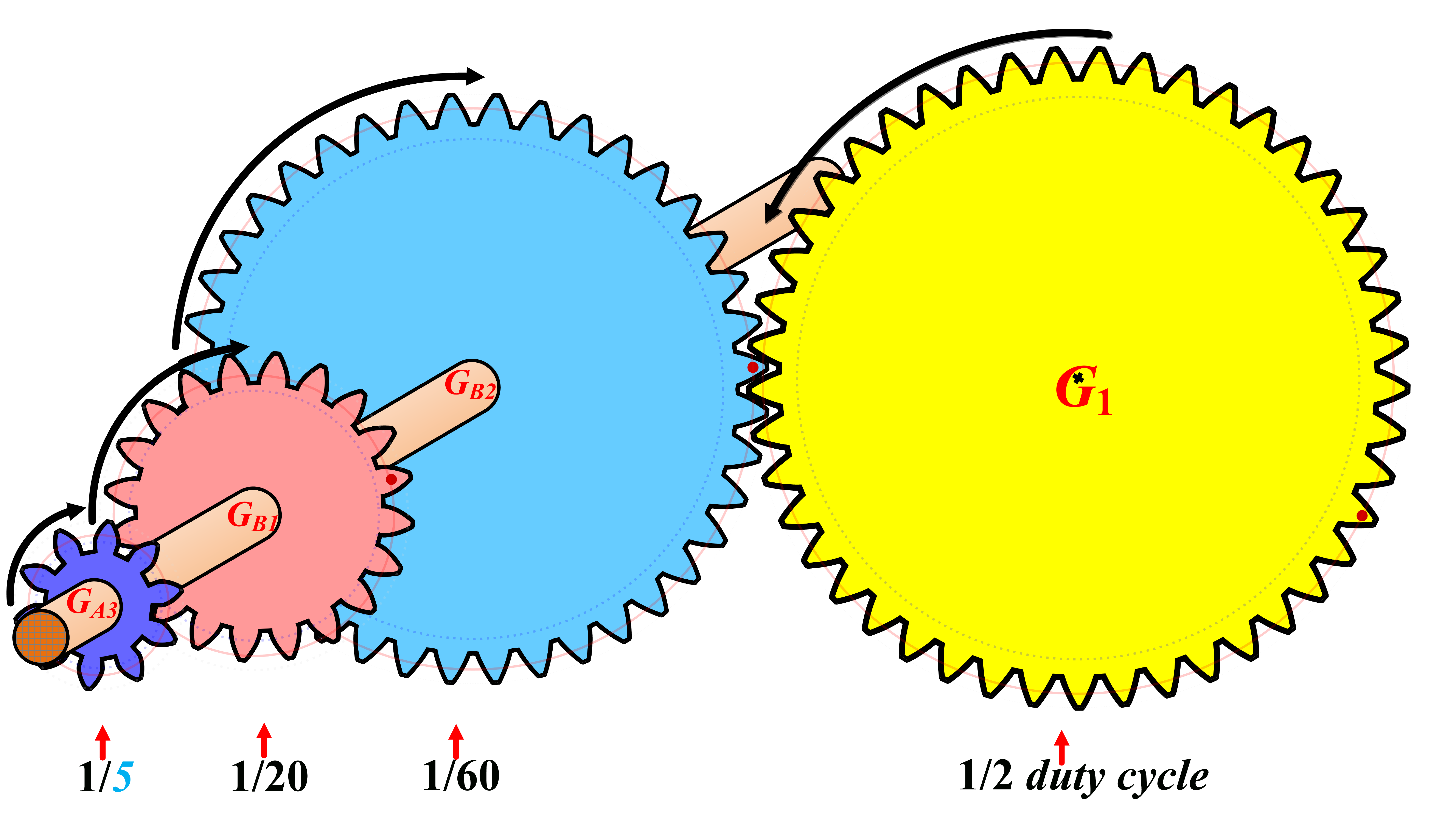}
\caption{Implementation of frequency multiplication with more gears.}\label{figdiss}
\end{figure}

\textbf{\textit{Another Inspiration.}} Motivated by the idea illustrated in Section \ref{bi}, if we use more than three oscillators to realize the frequency multiplication, a clock jitter would be produced. Although this operation to some extent could work, the final clock signal might be a little bias,  which uncovers another problem of coupling efficiency. Take a $1/60$ as an example. Since $60=3 \times 4 \times 5$, four oscillators are required, representing the duty cycle of $1/2$, $1/3$, $1/4$, and $1/5$, respectively. As illustrated in Fig. \ref{figdiss}, with the approach similar to Method 1, a single phase signal of $1/5$ duty cycle controls the whole transference of $1/4$ duty cycle. Then this controlling stream is towards $1/3$ and finally to $1/2$. Two phase signals are in demand of $1/3$ duty cycle to manipulate the rotation of $G_{1}$, which represents $1/2$ duty cycle.

\begin{figure}[htbp]
\centering
\subfigure[The produced signals for all gears.]{
\label{fig17:a}
\includegraphics[width=0.75\linewidth]{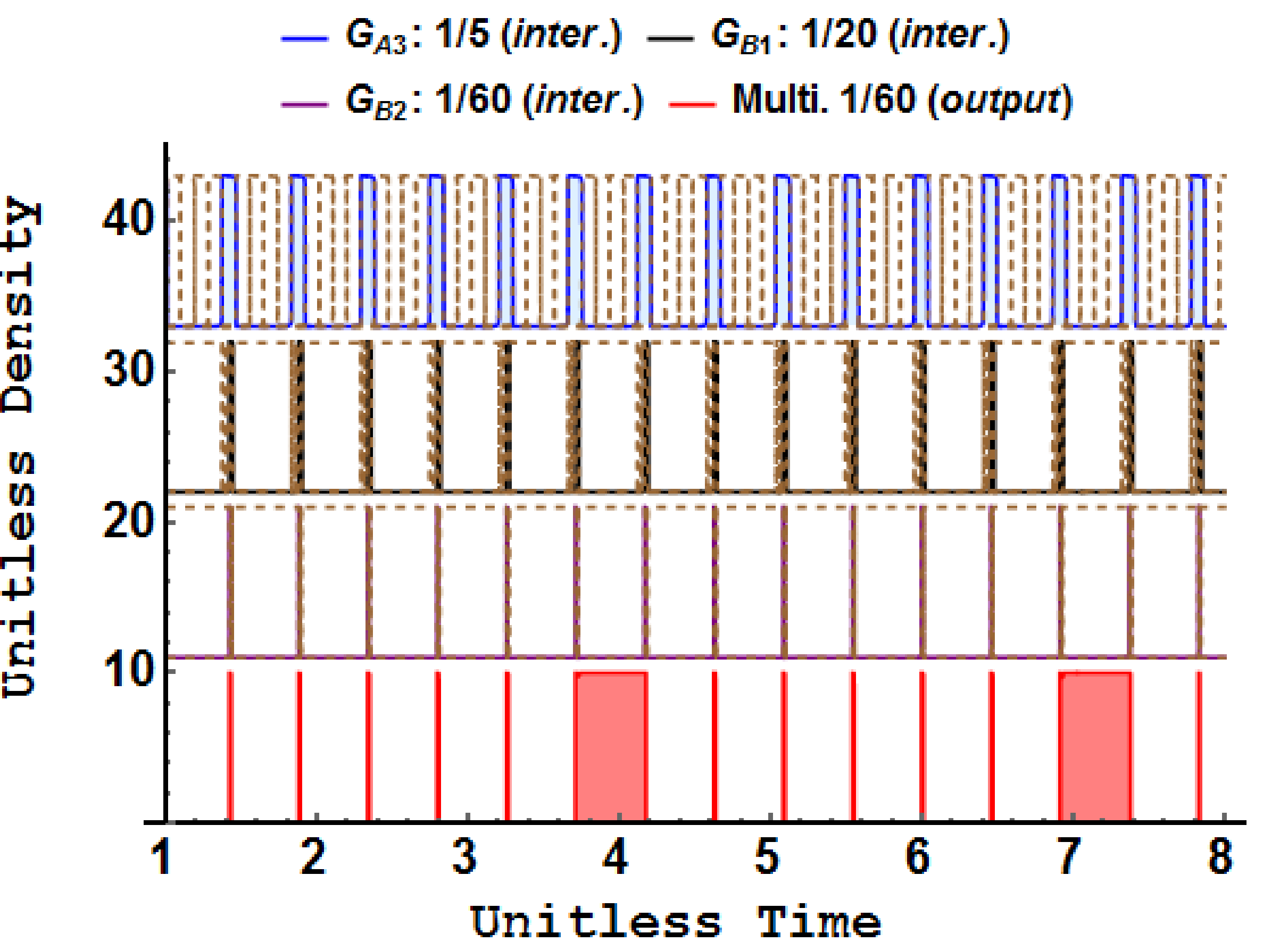}}
\subfigure[An enlarged version of all signals.]{
\label{fig17:b}
\includegraphics[width=0.75\linewidth]{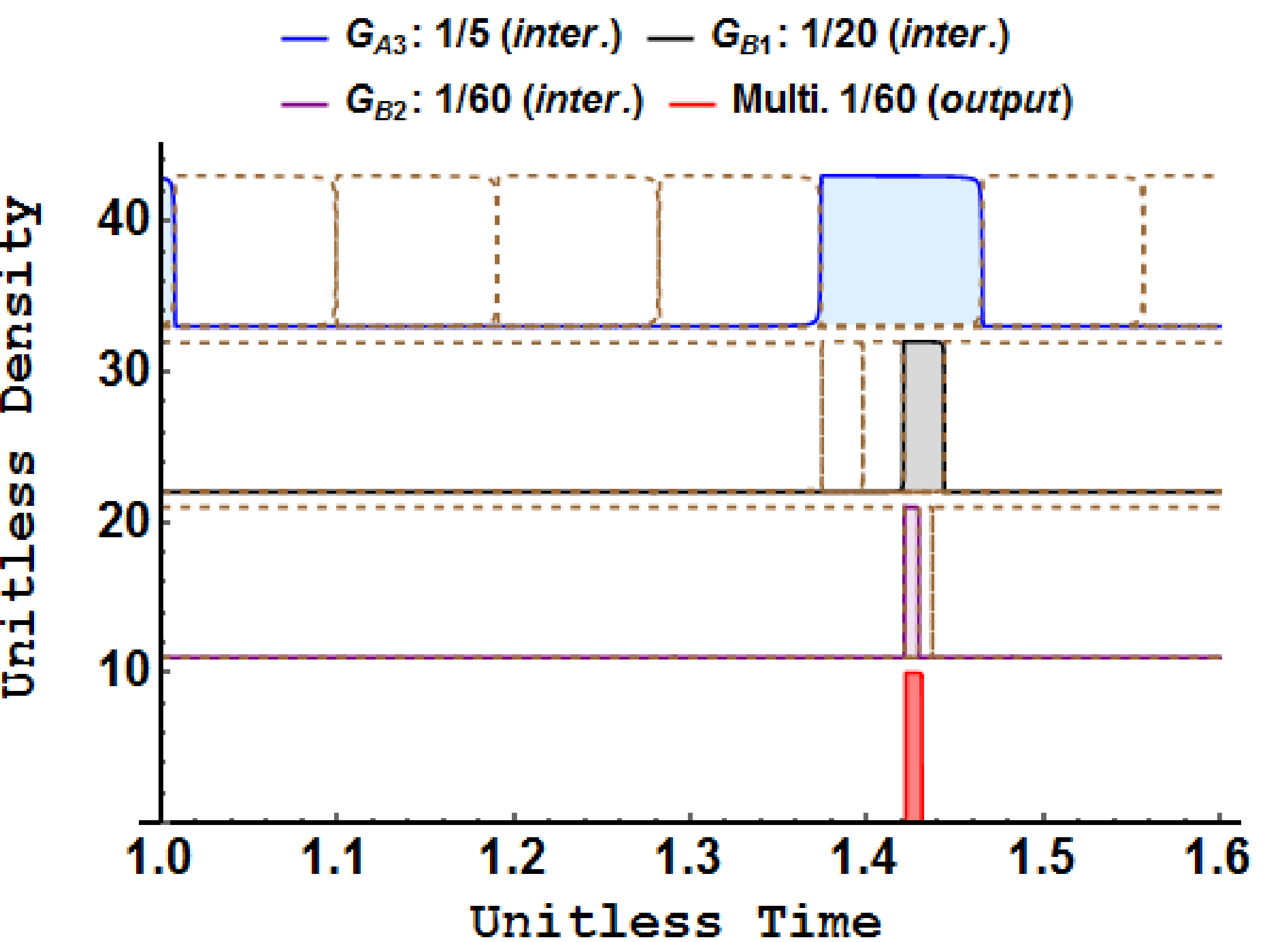}}
%\subfigure[The output of gear system.]{
%\label{fig17:a}
%\includegraphics[width=0.48\linewidth]{w121.pdf}}
%\subfigure[Simulation of single $G_{2}$.]{
%\label{fig17:b}
%\includegraphics[width=0.48\linewidth]{hh13.pdf}}
%\subfigure[Simulation of single $G_{6}$.]{
%\label{fig17:c}
%\includegraphics[width=0.48\linewidth]{w14.pdf}}
%\subfigure[Simulation of single $G_{4}$.]{
%\label{fig17:d}
%\includegraphics[width=0.48\linewidth]{w15.pdf}}
%\subfigure[Enlargement of the output signal.]{
%\label{fig17:e}
%\includegraphics[width=0.48\linewidth]{w12.pdf}}
%\subfigure[Enlargement of single $G_{2}$.]{
%\label{fig17:f}
%\includegraphics[width=0.48\linewidth]{w131.pdf}}
\caption{Results for frequency multiplication of $1/60$ with more gears.}\label{fig17}
\end{figure}

\textbf{\textit{Rate Constant Adjustment Scheme.}} Adopting the methods similar to Algorithm \ref{arg2}, no change exists in the rate constant of CRNs for the $5$-phase oscillator of $G_{A3}$. The rotation speed of $1/4$ duty cycle should be slowed down for its whole physical transferring period is shorter than $1/5$, and the rate constant of $1/3$ should be much faster than before. With the changed rate constant of main power reactions (duty cycle of $1/4$: $57$, $1/3$: $1100$), the corresponding simulation results are shown in Fig. \ref{fig17}.

\textbf{\textit{Analysis.}} From Fig. \ref{fig17}, simulation results, especially the top three curves, show the gears of $1/3$, $1/4$, and $1/5$ operate well. However in practical operations, gears shown in Fig. \ref{figdiss} might produce a flawed final frequency multiplication in the long run. This gear really works well as shown in the enlarged version of the first figure. If observe carefully, we can find that a blue-colored single phase of fundamental frequency for $1/5$ duty cycle is divided into four pieces. And this gray-colored single phase of $1/4$ duty cycle is segmented into three pieces. The bottom red curve is the final frequency multiplication of $1/60$ duty cycle, which is produced through a mesh between $G_{B2}$ and $G_1$. Note that this bottom red curve is actually a little bigger than the above purple one, and this kind of clock skew can be explained by the too long stage of meshing or coupling, since the standard single phase existing time is segmented again and again. Therefore, coupling efficiency in our gear model is also a problem worthy of more attention. Roughly, this tiny error of a little extended frequency multiplication for $1/60$ duty cycle might be omitted if the requirement is not strict. To our knowledge, we cannot completely solve the issue resulting from too many segmentations of \(\hat \tau _{1{\rm{/}}N}\) at this time. In our future work, we would like to handle this problem, as well as offer more quantitative analysis of our gear model in the aspect of information transport and clock skew. More experimental works with real DNA strand displacement reactions will also be offered.

\section*{Acknowledgment}
First, we would like to thank Dr. David Soloveichik for his kind help. We also would like to thank Editor-in-Chief, the associate editor, and the reviewers for their time and efforts to review this paper.

%, and for very constructive comments which helped to improve the quality and presentation of the paper.

%This work is supported in part by International Science \& Technology Cooperation Program of China under grant 2014DFA11640, Jiangsu Provincial Natural Science Foundation under grant BK20140636, Intel Collaborative Research Institute for Mobile Networking \& Computing, the Fundamental Research Funds for the Central Universities under grant 3204004202, the Research Fund of National Mobile Communications Research Laboratory, Southeast University under grant 2014B02, and the Project Sponsored by the Scientific Research Foundation for the Returned Overseas Chinese Scholars of State Education Ministry.

\ifCLASSOPTIONcaptionsoff
  \newpage
\fi
% ==================================================
\footnotesize
\bibliographystyle{IEEEtran}
\flushend
%% argument is your BibTeX string definitions and bibliography database(s)
\bibliography{IEEEabrv,mybib}

% that's all folks
\end{document}